\documentclass[preprint]{elsarticle}
\usepackage{graphicx,url,hyperref,microtype,subcaption,graphbox,amsmath}
\usepackage{latexsym,courier,upgreek,xcolor,rotating,ulem}
\usepackage{amssymb}
\usepackage{array}
\usepackage{paralist}
\usepackage[onehalfspacing]{setspace}
\hypersetup{colorlinks,citecolor=blue,linkcolor=red,urlcolor=green}
\usepackage[displaymath,mathlines]{lineno}
\usepackage[margin=2cm]{geometry}
\modulolinenumbers[5]
\biboptions{numbers,sort&compress}

\begin{document}

\begin{frontmatter}
  
  \title{Efficient fluid extraction through hydraulic fracture in capillary fiber bundle model}
  
  \author{Anjali Vajigi}
  \ead{p20240054@hyderabad.bits-pilani.ac.in}
  \author{Subhadeep Roy}
  \ead{subhadeep.r@hyderabad.bits-pilani.ac.in}
  \address{Dept. of Physics, Birla Institute of Technology \& Science Pilani, Hyderabad Campus, Secunderabad, Telangana 500078.}

\begin{abstract}
We have simulated a one dimensional capillary fiber bundle model with fracking events while acted between a pressure gradient across the system. The hydraulic fractures are incorporated through a decreasing nature of capillary thresholds for each tube that replicates an increment in pore spaces due to fracking. An increment in flow rate is evident through the evolved rheology we observe in our study. Analytical approaches for certain limits are adopted to understand the rheology which matches well with the numerical results. The overall hydraulic power increases with pressure gradient as well as with the percentage decrease in capillary threshold due to a single event, defines as the fracking amplitude. This combined with the early onset of linear Darcy flow increases the quality of the fluid extraction. We successfully point towards an optimum pressure gradient at which the fracking events are most effective - maximum change in fluid extracting with a maximum rate. We observed that it is possible to extract the information regarding the change from non-linear to Darcy flow due to fracking as well as the optimum pressure for fluid extraction through local flow profile, something which in much superior from the point of view of computational cost. The former is done by correlating the maximum fluctuation in local flow profile to the onset of Darcy flow. The later is done through the relative change in Shannon entropy with respect to the fracking amplitude that points towards the pressure associated with the maximum fluid extraction criterion. 
\end{abstract}


\end{frontmatter}



\section{Introduction}
Hydraulic fracturing, often referred  as fracking, has revolutionized the energy sector by enabling the extraction of hydrocarbons from unconventional reservoirs such as shale formations. As global energy demand continues to grow, driven by industrialization and urbanization, hydraulic fracturing has emerged as a key technology for unlocking resources that were previously inaccessible or uneconomical to exploit \cite{stat_rev_2022}. This technique, which involves injecting high-pressure fluids into subsurface rock formations to create fractures, has significantly increased oil and natural gas production \cite{us_energy_2022,wang2014}. Despite its benefits, hydraulic fracturing has raised significant environmental and societal concerns. Issues such as groundwater contamination, induced seismicity, and high water usage have been widely debated, leading to regulatory scrutiny and public opposition in some regions \cite{vidic2013,zoback2012}. These challenges highlight the need for continued research and innovation to minimize environmental impacts while maintaining operational efficiency. For example, recent studies have explored alternative fracturing fluids, improved wastewater treatment methods, and real-time monitoring systems to mitigate potential risks \cite{king2012,jiang2014}.

Understanding and optimizing hydraulic fracturing processes rely heavily on numerical and statistical modeling, which have become essential tools in both research and industry. Numerical models provide a detailed representation of the physical mechanisms involved in hydraulic fracture propagation, including fluid flow, rock deformation, and fracture mechanics \cite{adachi2007}. These models are often developed using finite element, finite difference, and boundary element methods, enabling simulations of complex interactions between rock formations, fracture networks, and fracturing fluids \cite{perkins1961,detournay2016}. For instance, fluid-driven fracture models have been extensively studied to predict fracture geometry, stress redistribution, and proppant transport \cite{geertsma1969}. Statistical modeling, on the other hand, is widely used to analyze field data and account for uncertainties in subsurface conditions and operational parameters. These models leverage large datasets from well completions and production outcomes to identify correlations and optimize treatment designs \cite{cipolla2011}. Machine learning techniques have further enhanced statistical modeling by enabling the analysis of high-dimensional datasets, improving predictions of fracture efficiency and production performance \cite{shahin2004}. For example, data-driven approaches have been employed to predict the productivity of hydraulically fractured wells based on inputs such as reservoir properties, injection parameters, and completion techniques \cite{kumar2017}. The integration of numerical and statistical models has led to significant advancements in fracture design and monitoring. Coupled models that combine geomechanics, fluid dynamics, and fracture mechanics offer detailed insights into fracture initiation and propagation under varying conditions \cite{peirce2008}. Meanwhile, probabilistic approaches allow for robust risk assessments, helping operators evaluate potential environmental impacts such as induced seismicity and fluid leakage \cite{maxwell2014}. Despite these advancements, challenges remain in scaling models from laboratory experiments to field conditions, as well as in validating model predictions with real-world data \cite{mcclure2014}.

In this paper, we have explored a statistical model, the capillary fiber bundle model, which tracks the propagation of interfaces through random media when a certain pressure gradient is applied across the system. The model due to its simplistic behavior allows us to tune different parameters to explore the role fracking. Moreover, this is the only model in the context of multi-phase flow in porous media that can be solved analytically for certain limits of disorder in capillary threshold. We explicitly showed that the fracking events leads to higher flow rate and different rheological behavior finally leading to better efficiency of fluid extraction - the ultimate goal of the present paper. The global transition point from non-linear to Darcy behavior along with the maximum extractivity point is finally matched with the local fluctuation in flow rate to bridge a connection between local and global parameters.   

In the next section we have provided a detailed description of capillary fiber bundle model and how the fracking events are implemented in it. This is followed by global rheological studies with increasing applied pressure and fracking amplitude and its prediction from local flow profile. Finally we have summarized our finding the conclusion section along with the direction towards future works.  

\begin{figure}[t]
\centerline{\includegraphics[width=0.8\textwidth,clip]{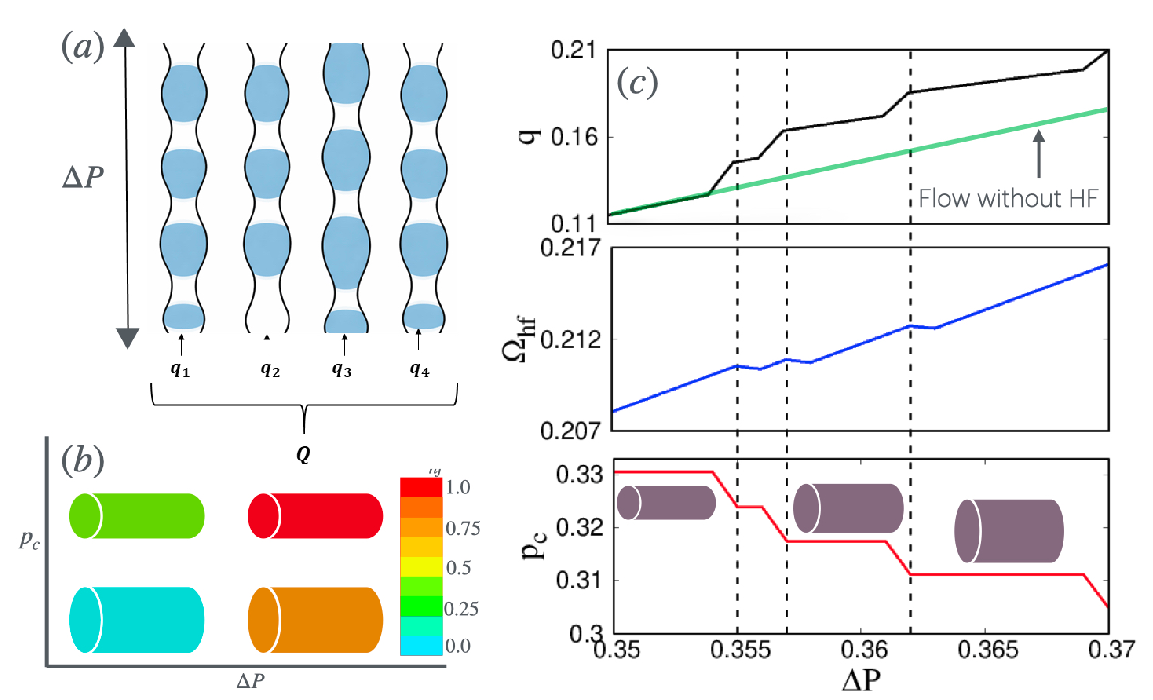}}
\caption{(a) The capillary fiber bundle model shown consists of $N$ parallel tubes acted between a same pressure gradient $\Delta P$. Each tube has a separate average diameter (not shown explicitly in the figure) around which the tubes keep a hour glass like shape of a certain periodicity. This combined with the number of interfaces inside a tube decides the capillary threshold pressure $p_c$. The dotted line is an imaginary cut through which the flow rate is calculated. \\ (b) Representation of the probability $\Omega_{hf}$ of hydraulic fracture with pressure gradient $\Delta P$ and $p_c$ (which is inverse of the link radius). $\Omega_{hf}$ is maximum (red) when both $\Delta P$ and $p_c$ is high. $\Omega_{hf}$ is moderate (orange) when $\Delta P$ is high but $p_c$ is low. $\Omega_{hf}$ is lower than moderate (green) when $\Delta P$ is low but $p_c$ is high as $\Delta P$ is a necessary condition for hydraulic fracture. $\Omega_{hf}$ is lowest (light blue) when both $\Delta P$ and $p_c$ are low. \\ (c) Variation of flowrate $q$ (upper), probability of hydraulic fracture $\Omega_{hf}$ (middle) and capillary pressure $p_c$ (lower) as a function of $\Delta P$ for a single tube. The straight line in the upper figure shows the flow of a tube without fracking. The schematic diagram of the tubes in lower figure shows increase in radius hence decrease in $p_c$ representing the fracking events.}
\label{fig0}
\end{figure}

\section{Model Description}
The capillary fiber bundle model (CFBM) is a bundle of  $N$ parallel capillary tubes each of length $L$ and cross sectional area $a$. The tubes are connected to a  pressure gradient $\Delta P$ (see figure \ref{fig0}a). The wetting and non-wetting fluid in each tube creates a bubble chain inside tube. If $L_w$ and $L_n$ are the lengths of wetting and non-wetting fluid respectively, the saturation can be written as: $S_w=L_w/L$ and $S_n=L_n/L$. Here, tubes are of sinusoidal shape to replicate simplified pore geometry, and the average radius is different for each tube. A bunch of long capillary tubes with varying radius can be seen as a series of many pores. Then the volume of wetting and non-wetting fluid in each tube will be $L_w a$ and $L_n a$ respectively. The total cross-sectional pore area of the total model is then $A_p=Na$. The spatial arrangement of the bubble interfaces inside the tube creates a random threshold pressure for each tube. In order to create flow in it one has to overcome this threshold.

The tubes can be seen as  connecting links between two pore bodies inside a porous media.  Due to the pressure drop the bubble interfaces move and capillary pressure for a bubble depends on the position of its interfaces. The contribution of all such bubble inside a tube along with the average radius and contact angle determines its  threshold pressure. In the recently explored work \cite{shs19}, the global flow rate is analytically determined as a function of increasing pressure drop. The pressure drop - flow rate rheology was explored by applying  fluctuation in the individual threshold pressure of tubes with varying disorder strength. It was observed that the existence of a lower cut off in the threshold pressure drop  highly effects the exponent in the non-linear region, while the results in the Darcy regime remains unaltered. For simplicity, we have not represented this threshold in terms of the fluid distribution. The contact angle is also assumed to be the same for all tubes without loosing any generality. The thresholds of each tube is drawn randomly from a certain distribution. The flow rate in certain tubes will be zero if the applied pressure gradient is less than its threshold value. Once the applied pressure exceeds its threshold, a certain tube starts flowing and starts contributing to the total flow rate. For analytical calculation and numerical simulation in CFBM, we have adopted the following threshold distribution: 
\begin{equation}\label{eq1}
\rho(p_c)=\left\{\begin{array}{ll}
                                0       & \mbox{, $p_c \le P_m$\;,}\\
                                \displaystyle\frac{1}{P_M-P_m} & \mbox{, $P_m < p_c \le P_M$\:,}\\
                                1       & \mbox{, $p_c > P_M$\;,}\\
              \end{array}
       \right.
\end{equation}
where $P_m$ and $P_M$ correspond to the lower and upper limits of $p_c$ respectively. Here each tube contains the same amount of fluid but has its own division of bubbles. 

The conventional CFBM is modified here by including the role of hydraulic fracture to see its effect in fluid extraction and on overall rheology during the multi-phase flow. This suggests that at any moment a tube can undergo a fracking event with a certain probability $\Omega_{hf}$ that will increase the radius of that link effectively decreasing its capillary threshold $p_c$. Ideally the probability of hydraulic fracture should depend on both the pressure gradient and the capillary threshold of a link. Figure \ref{fig0}(b) shows a schematic representation of this. The hydraulic fracture is most probable for a narrow pore under extreme pressure, in other words high $\Delta P$ and high $p_c$. $\Omega_{hf}$ is highest (red tube on top right) in this limit. Since $\Delta P$ is the driving force for the fracking event there will be less events when $\Delta P$ is low (green tube on top left) even for narrow pores ($p_c$ is high). On the other hand, a high pressure can still cause fracking for broad pores and $\Omega_{hf}$ for this case is kept higher (orange tube at right bottom) than the previous one. Finally, $\Omega_{hf}$ has a least value (light blue tube at the bottom left) for broad pores under low pressure when both $\Delta P$ and $p_c$ are low.          

This has been summarized as follows: 
\begin{align}\label{eq2}
\Omega_{hf} = \alpha_{\Delta P}\Delta P + \alpha_{p_c}p_c
\end{align}
The asymmetry in increasing $\Delta P$ vs increasing $p_c$ is established by considering $\alpha_{\Delta P}>\alpha_{p_c}$. In some sense $\alpha_{\Delta P}$ shows the strength of viscous force against the toughness of the body that contains the fluid. $\alpha_{p_c}$, on the other hand, shows the same interplay but with capillary pressure instead. Each tube is associated with a certain probability $\Omega_{hf}$ depending on its capillary pressure $p_c$ and the pressure gradient $\Delta P$ across the tube. A fracking event is carried on if $\Omega_{hf}$ is greater than a random value chosen from an uniform distribution $[0,1]$. After fracking, the radius of the tube increases, denoted by a reduction capillary pressure from $p_c$ to $p_c^{\prime}=(1-k/100)p_c$. The term $1-k/100$ accounts for the fact that the fracking event increases the radius from $r$ to $r+\Delta r$. Here, fracking amplitude $k$ represents the material strength. For weaker materials $k$ will be high suggesting larger increase in pore radius and hence larger decrease in capillary pressure. Smaller $k$, on the other hand, models a stronger material that can withstand the effect of high pressure and allows smaller change in pore radius. The vertical dotted line in figure \ref{fig0}(c) shows the fracking events where the radius of a tube is increased irreversibly. After this the flow rate for that $\Delta P$ is calculated. Next the pressure gradient is increased leading to new fracking events and flow rate $Q$ for that $\Delta P$. The change in a certain tube is represented in figure \ref{fig0}(c). The increase in radius of the tube is shown through the schematic diagram of the tubes that gets broaden after each hydraulic fracture event. This leads to sudden drop in $p_c$ of that tube. At the same point of fracking $\Omega_{hf}$ is also observed to decrease momentarily before increasing once again due to increasing $\Delta P$. Finally, the flow rate shows sudden jumps at the points where the hydraulic fracture events take place. The green straight line shows the flow rate in absence of the fracking events, which is lesser than the flow rate with hydraulic fracture. This leads to a modified rheology of the system with fracking events. The total flow rate in above case can be represented by the average over the ensemble of capillary tubes where the flow rate of each tube is obtained from the individual distribution of fluids. One assumption that has been taken here is a certain tube attains steady state immediately after the fracking enabling us to assume the flow rate $q \sim \sqrt{\Delta P^2-p_c^{{\prime}^2}}$ immediately after the event without any time delay.   

\section{Results and Discussion}
We have numerically studied a capillary fiber bundle model (CFBM) of system size $10^5$ and with $10^4$ configurations. The threshold pressure values are chosen from equation \ref{eq1} while the external pressure is increased from 0 to 10 (in arbitrary units). At a certain pressure the whole bundle is scanned to locate the fracking events by comparing a random number from the range 0 to 1 with probability $\Omega_{hf}$ (see equation \ref{eq2}). 

\begin{figure}[ht]
\centerline{\includegraphics[width=0.5\textwidth,clip]{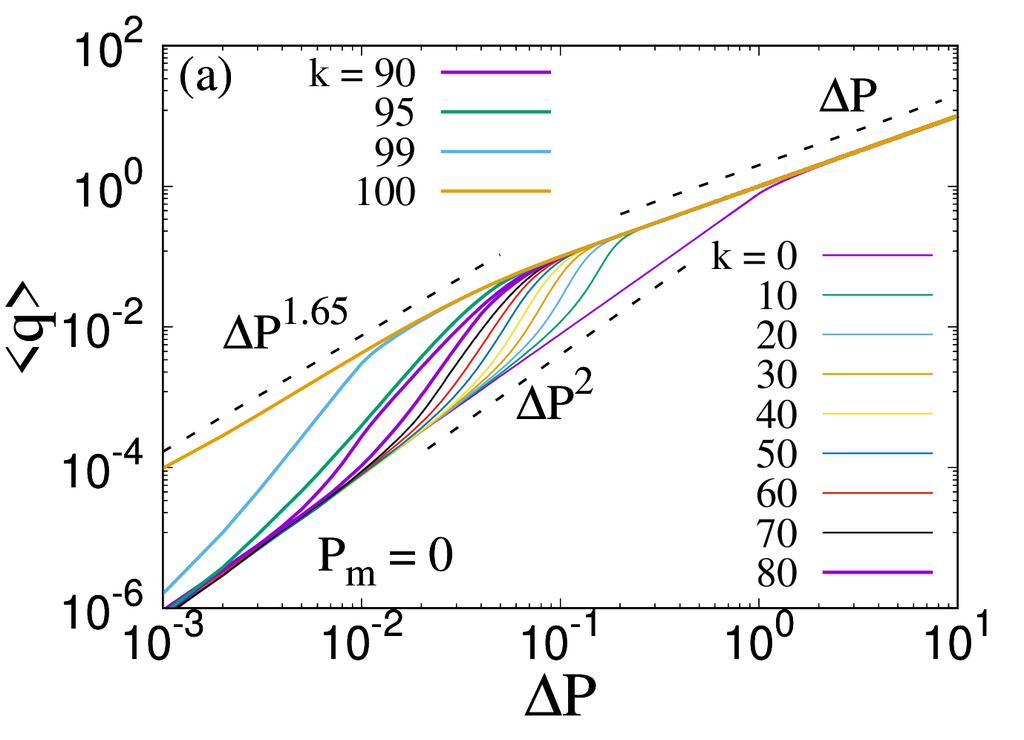}}
\centerline{\includegraphics[width=0.5\textwidth,clip]{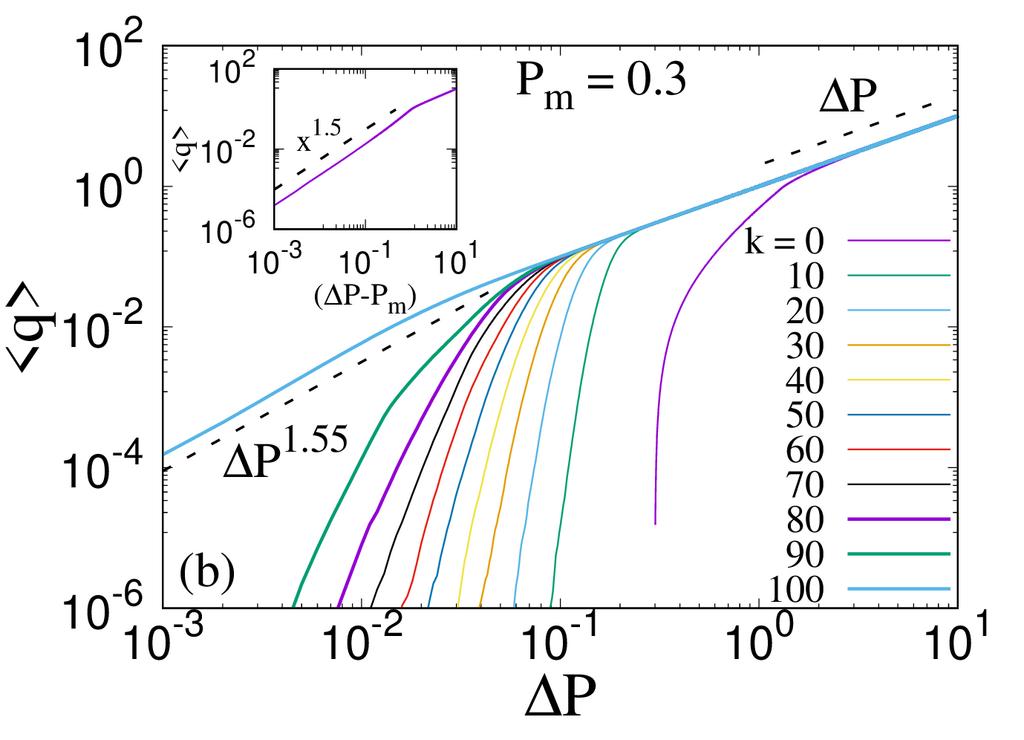}}
\caption{Effective rheology with fracking events for (a) $P_m=0$ and (b) $P_m=0.3$ with varying $k$ - the percentage with which $p_c$ reduces. For $P_m=0$, the behavior at low $\Delta P$ deviates from $Q \sim \Delta P^{2}$ and finally shows the scaling $Q \sim \Delta P^{1.65}$ when $k=100$. At high $\Delta P$, on the other hand, the rheology becomes Darcy like linear. For $P_m=0.3$, on the other hand, we do not observe any scale free rheology (though flow rate increases with $k$) except for $k=100$ where we see $Q \sim \Delta P^{1.55}$ again when pressure gradient is less. At high $\Delta P$, in this case as well, we observe the linear Darcy flow.}
\label{fig3}
\end{figure}

\subsection{The Steady state flow}
Figure \ref{fig3} shows the effective rheology of the CFBM in presence of the hydraulic fracture events when the pressure gradient $\Delta P$ across the system is increased from $10^{-3}$ to $10$. We have tuned the parameter $k$, the percentage of decrease in capillary pressure due to a single fracking event, from 0 (suggesting no hydraulic fracture) to 100 (severe fracking where $p_c$ goes to zero in a single event). Figure \ref{fig3} shows the results for $P_m=0$ (figure \ref{fig3}a) and $P_m=0.3$ (figure \ref{fig3}b). For $k=0$, we observe the convention rheology - $\langle q \rangle \sim \Delta P^2$ for $P_m=0.0$ and $\langle q \rangle \sim (\Delta P-P_m)^{1.5}$ for $P_m=0.3$ (see inset of figure \ref{fig3}b) when pressure gradient is sufficiently low. As we increase $k$, allowing the hydraulic fracture events, $\langle q \rangle$ increases for both $P_m=0$ and $P_m>0$, but shows a scale-free rheology, $\langle q \rangle \sim \Delta P^2$, only for the former case in the limit $\Delta P\rightarrow P_m$. This suggests a higher flow rate due to the fracking events which we will discuss in details later. When $k=100$, we observe a scale free rheology with exponent closer to $1.55$ or $1.65$ depending on whether the lower cutoff is present or not. This happens, most probably because of the fact that, $k=100$ suggests an 100\% of reduction in capillary pressure and treat that link to be a free path when encounter a fracking event. threshold of any fiber that experience the fracking goes to zero, effectively making the threshold distribution from zero even if a finite cut-off $P_m$ is there. Below we will try to validate the numerical results through analytical calculations.  

\subsubsection{Analytical results for steady-state rheology}
The capillary fiber bundle model, being a simple model, should be able to provide some analytical insight to the rheological behavior. The volumetric flow rate in a capillary tube of length $L$ is given by \cite{shbk13} 
\begin{align}\label{eq3}
q=-\displaystyle\frac{a^2}{8\pi\mu_{av}L}\Theta(\Delta P-p_c)(\Delta P-p_c)
\end{align}
where $\Delta P$ is the pressure drop across the capillary tube, $a$ is the radius of the link, $p_c$ the overall capillary force due to all interfaces within a certain capillary tube, and $\mu_{av}$ is the effective viscosity. $\Theta(\Delta P-p_c)$ is the Heaviside function which is 1 for $\Delta P>p_c$ and zero otherwise. Then in the steady state, the time-averaged flow rate is given by, 
\begin{align}\label{eq4}
\bar{q}=-\displaystyle\frac{a^2}{8\pi\mu_{av}L}sgn(\Delta P)\Theta(\Delta P-p_c)\sqrt{\Delta P^2-p_c^2}
\end{align}
Then at a constant pressure gradient $\Delta P$, the normalized volumetric flow rate for the whole bundle with $N$ tubes is given by, 
\begin{align}\label{eq5}
\langle q \rangle =-\displaystyle\frac{a^2}{8N\pi\mu_{av}L}\displaystyle\int_{P_m}^{max(\Delta P, P_M)}\sqrt{\Delta P^2-p_c(t)^2} \ \rho(p_c(t))dp_c
\end{align}
where $p_c(t)$ is the time dependent capillary pressure that changes due to the fracking events depending on $\Omega_{hf}$ and $k$. The upper limit of the integration depends on whether the applied pressure $\Delta P$ is great than or less than maximum possible threshold $P_M$. 

We will discuss the analytical results for $k<100$ (Case I) first and then for $k=100$ (Case II),  the latter being the limit where the capillary threshold is completely removed due to a single fracking event. For each case, we will discuss the role of lower cutoff of threshold distribution as well as the position of the pressure gradient $\Delta P$. 

Whenever a tube goes through a fracking event, the capillary pressure reduces by $k\%$, producing the following $p_c$ after $n$ fracking events, 
\begin{align}\label{eq6}
p_c^n(t)=\left[1-\displaystyle\frac{k}{100}\right]^np_c
\end{align}
This decrease happens for all tubes including the upper limit. The integration showing global volumetric flow rate will then be
\begin{align}\label{eq7}
\langle q \rangle =-\displaystyle\frac{a^2}{8N\pi\mu_{av}L}\displaystyle\int_{0}^{min(\Delta P, P_M^{\prime})}&\sqrt{\Delta P^2-\left[\left(1-\displaystyle\frac{k}{100}\right)^np_c\right]^2} \nonumber \\ &\rho(p_c(t))dp_c
\end{align}
where $P_M^{\prime}=P_M\left(1-\displaystyle\frac{k}{100}\right)^n$. \\

\textbf{Case I(a): $k<100$, $P_m=0$, $\Delta P<P_M$ - } In the limit, $\Delta P<P_M$, we have,
\begin{align}\label{eq8}
\langle q \rangle =&-\displaystyle\frac{a^2}{8N\pi\mu_{av}L}\displaystyle\frac{1}{P_M\left(1-\displaystyle\frac{k}{100}\right)^n} \displaystyle\int_{0}^{\Delta P}\sqrt{\Delta P^2-\left[\left(1-\displaystyle\frac{k}{100}\right)^np_c\right]^2} dp_c \nonumber \\
=&-\displaystyle\frac{a^2}{8N\pi\mu_{av}L}\displaystyle\int_{0}^{\Delta P}\sqrt{\displaystyle\frac{\Delta P^2}{P_M^2\left(1-\displaystyle\frac{k}{100}\right)^{2n}}-\left(\displaystyle\frac{p_c}{P_M}\right)^2} dp_c
\end{align}
where $\rho(p_c(t))=1/P_M^{\prime}=1/[P_M(1-k/100)^n]$. Only changes close to $P_M$ is considered for low $\Delta P$ as $\Omega_{hf}$ in this limit will mostly depend on $p_c$, which has higher value closer to $P_M$. With the transformation $z=\displaystyle\frac{\Delta P}{P_M(1-\frac{k}{100})^n}$ and $x=p_c/P_M$, we can rewrite the above integration as, 
\begin{align}\label{eq9}
\langle q \rangle =-\displaystyle\frac{a^2P_M}{8N\pi\mu_{av}L}\displaystyle\int_{0}^{\Delta P/P_M}\sqrt{z^2-x^2} \ dx
\end{align}
With another transformation $x=z\sin\theta$, we have,
\begin{align}\label{eq10}
\langle q \rangle =-\displaystyle\frac{a^2P_Mz^2}{8N\pi\mu_{av}L}\displaystyle\int_{0}^{\sin^{-1}(\Delta P/zP_M)}\cos^2\theta \ d\theta
\end{align}
This gives the final expression for the volumetric flow rate to be, 
\begin{align}\label{eq11}
\langle q \rangle =-\displaystyle\frac{a^2}{8N\pi\mu_{av}L}\displaystyle\frac{\Delta P^2}{2P_M}\displaystyle\frac{1}{\left(1-\displaystyle\frac{k}{100}\right)^{2n}}&\Bigg[\arcsin\left(1-\displaystyle\frac{k}{100}\right)^n \nonumber \\ &-\left(1-\displaystyle\frac{k}{100}\right)^n\sqrt{1-\left(1-\frac{k}{100}\right)^{2n}}\Bigg]
\end{align}
Equation \ref{eq11} shows a non-linear increase in global flow rate: $\langle q \rangle \sim \Delta P^2$. This behavior is observed numerically as well in figure \ref{fig3}(a) for $P_m=0$ and for low $\Delta P$. For $k>0 (\ne 100)$, the $\langle q \rangle-\Delta P$ behavior is quadratic for low $\Delta P$, gradually deviates from it with increasing $\Delta P$ and finally meets the linear Darcy behavior. The rheology for high $\Delta P$ limit is discussed next. \\

\textbf{Case I(b): $k<100$, $P_m=0$, $\Delta P>P_M$ -} For $\Delta P>P_M$, equation \ref{eq8} becomes, 
\begin{align}\label{eq12}
\langle q \rangle=&-\displaystyle\frac{a^2}{8N\pi\mu_{av}L}\displaystyle\frac{1}{{P_M}\left(1-\dfrac{k}{100}\right)^n} \displaystyle\int_{0}^{{P_M}\left(1-\dfrac{k}{100}\right)^n}\sqrt{\Delta P^2-\left[\left(1-\displaystyle\frac{k}{100}\right)^np_c\right]^2} dp_c
\end{align}
This will lead to the result below: 
\begin{align}\label{eq13}
\langle q \rangle = -\displaystyle\frac{a^2}{8N\pi\mu_{av}L}\displaystyle\frac{\Delta P^2}{2P_M\left(1-\frac{k}{100}\right)^{2n}} &\Bigg[\arcsin{\left(\dfrac{P_M\left(1-\frac{k}{100}\right)^{2n}}{\Delta P}\right)} \nonumber \\
&+\dfrac{P_M\left(1-\frac{k}{100}\right)^{2n}}{\Delta P}\sqrt{1-\left(\dfrac{P_M\left(1-\frac{k}{100}\right)^{2n}}{\Delta P}\right)^2}\Bigg] 
\end{align}
In the limit, $\Delta P>>P_M$, the above flow rate takes the form, 
\begin{align}\label{eq14}
\langle q \rangle = \displaystyle\frac{a^2}{8N\pi\mu_{av}L}\Delta P
\end{align}
In the limit of very high pressure, we observe the linear Darcy behavior as expected. This also matches with what we observe numerically (see figure \ref{fig3}a). \\

\textbf{Case I(c): $k<100$, $P_m>0$, $\Delta P<P_M$ -} The dynamics between global flow rate $Q$ and pressure gradient becomes more complicated when we include a lower cutoff to the distribution of capillary threshold. This is due to the fact that now both $P_m$ and $P_M$ reduces since fibers closer to upper cutoff as well as lower cutoff are prone to hydraulic fracture, though the chances of the latter will be higher. If we assume the weakest and the strongest capillary tube goes through $n$ fracking events, one can write the cutoffs as follows: 
\begin{align}
&P_M^{\prime}=P_M\left(1-\frac{k}{100}\right)^{n} \ \ \ \ \text{and} \nonumber \\
&P_m^{\prime}=P_m\left(1-\frac{k}{100}\right)^{n}
\end{align}
For now, we wont need $P_M^{\prime}$ as we will integrate up to $\Delta P$ to find the global flow rate:
\begin{align}
 \langle q \rangle =  -\displaystyle\frac{a^2}{8N\pi\mu_{av}L}\dfrac{1}{\left(P_M-{P_m}\right)\left(1-\dfrac{k}{100}\right)^n}\displaystyle\int_{P_m\left(1-\frac{k}{100}\right)^n}^{\Delta P}\sqrt{\Delta P^2-\left[{p_c}\left(1-\dfrac{k}{100}\right)^n\right]^2} dp_c
\end{align}
In the limit $\Delta P \rightarrow P_m$ and with fracking events, individual capillary thresholds also approaches towards $P_m$. The integration is performed with the following transformation, to get the final expression for $\langle q \rangle$: 
\begin{align}
\langle q \rangle = -\displaystyle\frac{a^2}{8N\pi\mu_{av}L}\frac{2}{3}\frac{1}{P_M-P_m} \sqrt{\frac{P_m}{\left(1-\frac{k}{100}\right)^n}}\left[\Delta P-P_m\left(1-\frac{k}{100}\right)^n\right]^\frac{3}{2}
\end{align}
For $k=0$, one obtains the conventional behavior: $\langle q \rangle \sim (\Delta P-P_m)^{3/2}$ \cite{shs19}. \\

\textbf{Case I(d): $k<100$, $P_m>0$, $\Delta P>P_M$ -} Similar to Case I(c), we can find the global flow rate here by carrying out the above integration $P_m^{\prime}$ to $P_M^{\prime}$.   
\begin{align}
 \langle q \rangle &=  -\displaystyle\frac{a^2}{8N\pi\mu_{av}L}\dfrac{1}{\left(P_M-{P_m}\right)\left(1-\dfrac{k}{100}\right)^n}\displaystyle\int_{P_m\left(1-\frac{k}{100}\right)^n}^{P_M\left(1-\frac{k}{100}\right)^n}\sqrt{\Delta P^2-\left({p_c}\left(1-\dfrac{k}{100}\right)^n\right)^2} dp_c
\end{align}
With proper substitution mentioned earlier, we get, in the limit $\Delta P>>P_M$, 

\begin{align}
\langle q \rangle = -\displaystyle\frac{a^2}{8N\pi\mu_{av}L}&\dfrac{\Delta P^2}{2\left(P_M-P_m\right) \left(1-\frac{k}{100}\right)^{2n}} \nonumber \Bigg[\left(\frac{P_M}{\Delta P}\left(1-\frac{k}{100}\right)^{2n}\right)- \left(\frac{P_m}{\Delta P}\left(1-\frac{k}{100}\right)^{2n}\right) \nonumber \\ &+ \dfrac{P_M}{\Delta P}\left(1-\frac{k}{100}\right)^{2n} -\dfrac{P_m}{\Delta P}\left(1-\frac{k}{100}\right)^{2n}\Bigg] 
\end{align}

Again, in the limit $\Delta P>>P_M$ above result shows the following Darcy like dependency
\begin{align}
\langle q \rangle = -\displaystyle\frac{a^2}{8N\pi\mu_{av}L}\Delta P
\end{align}

\textbf{Case II: $k=100$ -} This is a special case where the threshold pressure goes to absolute zero as soon as one tube meets a fracking event. This in turn leaves the rest of the system same. After $n$ fracking events for a bundle consisting $N$ fibers initially, the behavior in this limit can be understood separately as two parts: (i) $n$ number of fibers with threshold pressure zero, plus (ii) rest $N-n$ fibers distributed uniformly but with a reduced width of the distribution $P_M^{\prime\prime}$ where $P_M^{\prime\prime}=P_M(\frac{N-n}{N})$.

For $P_m=0$, the global flow rate, in the limit $\Delta P<P_M^{\prime\prime}$ will then be,
\begin{align}\label{eq15}
\langle q \rangle &= -\displaystyle\frac{a^2}{8N\pi\mu_{av}L}\left[n.\Delta P+\displaystyle\frac{1}{P_M^{\prime\prime}}\displaystyle\int_{0}^{\Delta P}\sqrt{\Delta P^2-p_c^2} \ dp_c\right]  \nonumber \\
  &= -\displaystyle\frac{a^2}{8N\pi\mu_{av}L}\left[n.\Delta P+ \displaystyle\frac{\pi}{4P_M^{\prime\prime}}\Delta P^2\right]
\end{align}
This includes both the linear and the non-linear term together. On the other hand, in the limit $\Delta P>P_M^{\prime\prime}$ will be, 
\begin{align}\label{eq16}
&\langle q \rangle = -\displaystyle\frac{a^2}{8N\pi\mu_{av}L}\left[n\Delta P+\displaystyle\frac{1}{P_M^{\prime\prime}}\displaystyle\int_{0}^{P_M^{\prime\prime}}\sqrt{\Delta P^2-p_c^2} \ dp_c\right]  \nonumber \\
&= -\displaystyle\frac{a^2}{8N\pi\mu_{av}L}\left[n\Delta P+ \displaystyle\frac{\Delta P^2}{2P_M^{\prime\prime}}\left(\displaystyle\frac{P_M^{\prime\prime}}{\Delta P}+\displaystyle\frac{P_M^{\prime\prime}}{\Delta P}\sqrt{1-\left(\displaystyle\frac{P_M^{\prime\prime}}{\Delta P}\right)^2}\right)\right]
\end{align}
This shows the linear Darcy behavior when $\Delta P>>P_M^{\prime\prime}$, 
\begin{align}\label{eq17}
\langle q \rangle &= -\displaystyle\frac{a^2}{8N\pi\mu_{av}L}(n+1)\Delta P
\end{align}

For $k=100$ and $P_m>0$ limit, above integration will be modified as: 
\begin{align}\label{eq18}
\langle q \rangle &= -\displaystyle\frac{a^2}{8N\pi\mu_{av}L}\left[n\Delta P+\displaystyle\frac{1}{P_M^{\prime\prime}-P_m}\displaystyle\int_{P_m}^{\Delta P}\sqrt{\Delta P^2-p_c^2} \ dp_c\right] \  \text{or} \nonumber \\
&= -\displaystyle\frac{a^2}{8N\pi\mu_{av}L}\left[n\Delta P+\displaystyle\frac{1}{P_M^{\prime\prime}-P_m}\displaystyle\int_{P_m}^{P_M^{\prime\prime}}\sqrt{\Delta P^2-p_c^2} \ dp_c\right]
\end{align}
depending on whether $\Delta P<P_M^{\prime\prime}$ or $>P_M^{\prime\prime}$ respectively. This will lead to a behavior $\langle q \rangle \sim \Delta P+(\Delta P-P_m)^{3/2}$ for the former case and $\langle q \rangle\sim\Delta P$ when $\Delta P>>P_M^{\prime\prime}$ for the latter. This does not agree with the numerical result where we observe $\langle q \rangle \sim \Delta P^{1.55}$ in the presence of lower cutoff and $\langle q \rangle \sim \Delta P^{1.65}$ when the lower cutoff is not there.

\subsubsection{Spatial distribution of kinetic energy}
As the dynamics we are discussing here are completely flow dominated, we wanted to explore participation number($\pi$), a quantity that tells us about the the distribution of kinetic energy among the flowing tubes. We believe the nature of the participation number will be able to shed more light on the steady state rheology. The distribution of energies in individual flow path will help us to decode the underlying complexities, as well as, one can clearly differentiate the flow regimes based  on the energies. The participation number for a complex flow is defined as
\begin{align}
\pi = N\sum_{i=1}^{N}\xi_i^2
\end{align}
where $\xi_i$ is the relative contribution of a single tube in terms of kinetic energy compared to the total kinetic energy of the system. $\xi_i$ is then expressed as
\begin{align}
\xi_i = \dfrac{k_i}{\sum_{i=1}^N k_i}
\end{align}
 $k_i$ being the kinetic energy generated by the $i^{th}$ link while the denominator is the total kinetic energy produced through $N$ flowing channels. 
\begin{figure}[ht]
\centerline{\includegraphics[width=0.5\textwidth,clip]{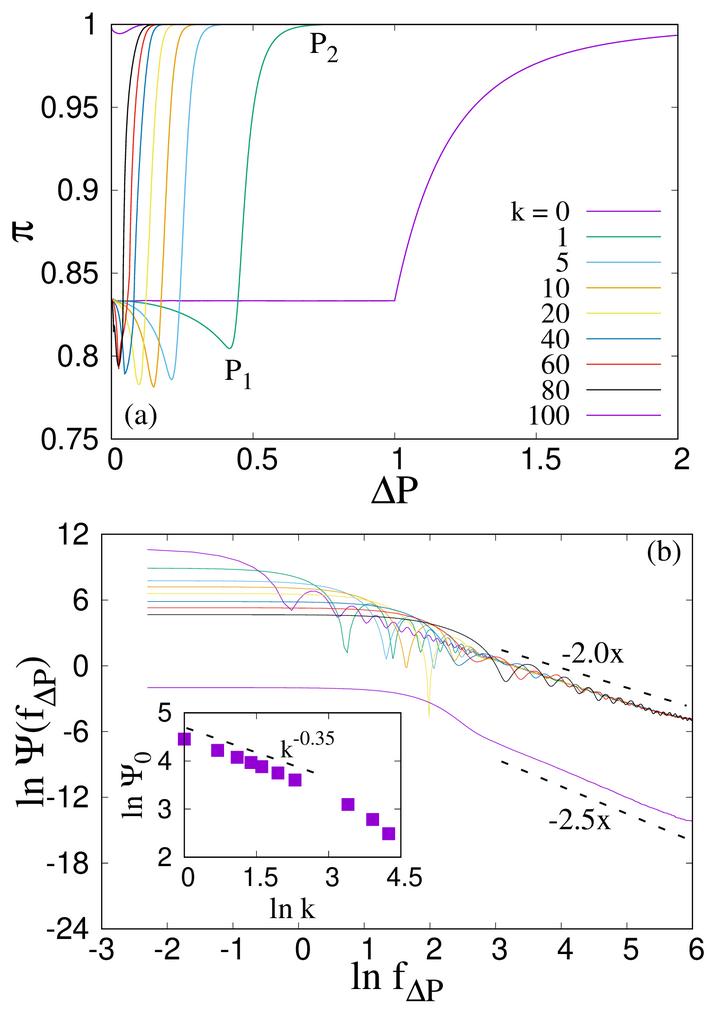}}
\caption{(a) Participation number $\pi$ as a function of $\Delta P$ for different $k$ ranging in between 0 and 100. $\pi$ reaches its lowest value at $P_1$, which later goes towards 1 at $P_2$ suggesting equipartition of energy. (b) The power spectrum $\Psi(f_{\Delta P})$ with frequency $f_{\Delta P}$ showing the following scaling: (i) $\Psi(f_{\Delta P})\sim f_{\Delta P}^{-2}$ for $k<100$, and (ii) $\Psi(f_{\Delta P})\sim f_{\Delta P}^{-5/2}$ for $k=100$. This shows a shift from red to black noise.}
\label{fig15}
\end{figure}
 The above ratio then expresses the fraction of kinetic energy that a tube shares compared to the cumulative kinetic energy of the system. Newtonian mechanics tells that the term $k_i$ includes the fluid velocity $v_i$ which again can be written as the ratio of flow rate and cross-sectional area of a tube. If we assume the fluid flowing through a tube of unit area and having a unit mass, then the kinetic energy will be related to the flow rate as follows: $k_i \propto v_i^2 \propto q_i^2$.

\begin{figure}[ht]
\centerline{\includegraphics[width=0.45\textwidth,clip]{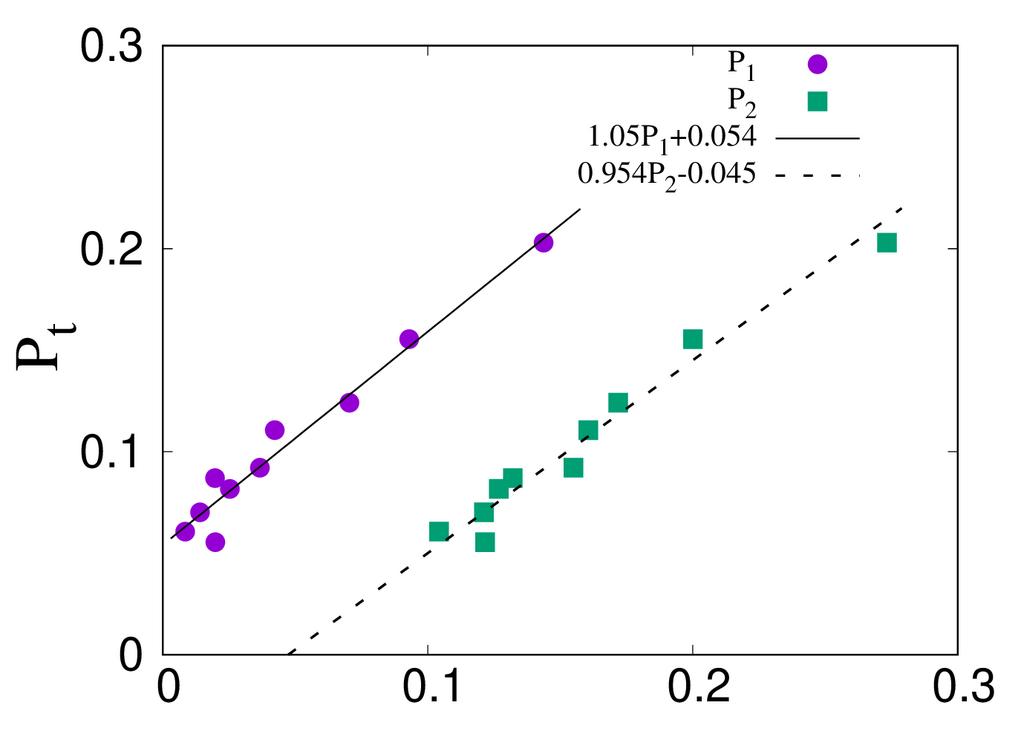}}
\caption{The figure shows the following correlation among the the onset of Darcy flow ($P_t$), the minimum point in participation number ($P_1$) and the scenario showing equipartition of energy ($P_2$): $P_t=1.05P_1+0.050$ and $P_t=0.95P_2-0.045$.}
\label{fig20}
\end{figure}

Figure \ref{fig15} shows how the participation number $\pi$ varies with applied pressure $\Delta P$ when the fracking amplitude $k$ is tuned. We observe, for $k>0$, $\pi$ starts from a value in between 0.8 and 0.85, decreases momentarily with $\Delta P$ and then increases towards 1 as we increase $\Delta P$. For $k=0$, the initial decrease is not observed and $\pi$ remains constant before it increases. $\pi=1$ suggests equipartition of energy. In this limit, since $\Delta P>>P_M$, the flow-rate in all channels is almost $\Delta P$ since this satisfies $\Delta P>>p_c$ as well. For higher $k$, we observe $\pi$ to reach 1 much early. This happens as for higher $k$, most of the tubes will become active and as we increase pressure the flow rate in each path becomes similar. We observe a minima in $\pi$ (say, at $\Delta P=P_1$) and finally increases again to reach 1 (say, at $\Delta P=P_2$) eventually when the pressure gradient is further increased. The correlation shown in figure \ref{fig20} shows $P_1$ coincides with the onset of Darcy flow $P_t$ very closely. 
This is understandable as a reduction in $\pi$ represents more fluctuation in the flow profile and the model goes away from the equipartition of energy. This will happen as $\Delta P$ gradually approaches $P_t$ and more tubes start to flow, contribution to the total fluctuation and hence a decreasing participation number. As $\Delta P$ crosses $P_t$, no more tubes are expected to open but since the gap between $\Delta P$ and the individual threshold values increases, all flow rates starts to become equivalent to $\Delta P$. The participation number $\pi$ again gradually increases in this region and eventually reaches 1.  
\begin{align}
    &P_t=1.05P_1+0.050    \nonumber \\
    &P_t=0.95P_2-0.045
\end{align}
Above equations explicitly shows the correlation involving $P_1$, $P_2$ and $P_t$. $P_1$ takes place just before the onset of Darcy flow and equipartition occurs after the model enters the Darcy region.  

To gain deeper insight into the transitions in the participation number, we analyze the participation number in the frequency domain by studying its power spectrum. The power spectrum of participation number shows signature of two types of behavior, 
\begin{align}
&\Psi(f_{\Delta P})\sim f_{\Delta P}^{-2} \ \ \ \text{for} \ \ \ k<100 \ \ \ \text{and} \nonumber \\
&\Psi(f_{\Delta P})\sim f_{\Delta P}^{-5/2} \ \ \ \text{for} \ \ \ k=100
\end{align}
In frequency space, the power spectrum showing scale free behavior $f^{-\beta}$ indicating colored noise based on the exponent $\beta$.
The $f_{\Delta P}^{-2}$ behavior is similar to the red noise, often observed in Brownian-like motion, random walk or diffusion process where individual events are uncorrelated from one another. The presence of red noise in porous media is already well studied which indicates towards self organized criticality in the system \cite{vr25}. Here,  as we focus on how an event of fracking can change the dynamics. We can clearly see  For $k=100$, the exponent jumps to 2.5 generally known as black noise \cite{schroeder2000} where the power spectrum exponent $\beta>2$,  which suggests slight increase in correlation, possibly due to the occurrence of larger avalanches as the fiber thresholds becomes zero. Such 2.5 exponents are often observed for correlated fracture processes, plastic deformation, earthquake-like stick-slip, fluid invasion in disordered media, etc. The inset of figure \ref{fig15}(b) shows the spectrum $\Psi_0$ at very high pressure gradient (towards zero frequency), where equipartition of energy takes place. We observe, $\Psi_0$ to decay in a scale-free manner, alteast at lower $k$ values as follows: $\Psi_0 \sim k^{-0.35}$. This decrease suggests that the global re-organization at large pressure is being suppressed as the fluctuation fades. This is true since with higher $k$, the threshold values comes closer making the re-organization less and less effective. 


\subsubsection{Prediction of steady-state rheology from local flow behavior}
\begin{figure}[ht]
\centerline{\includegraphics[width=0.7\textwidth,clip]{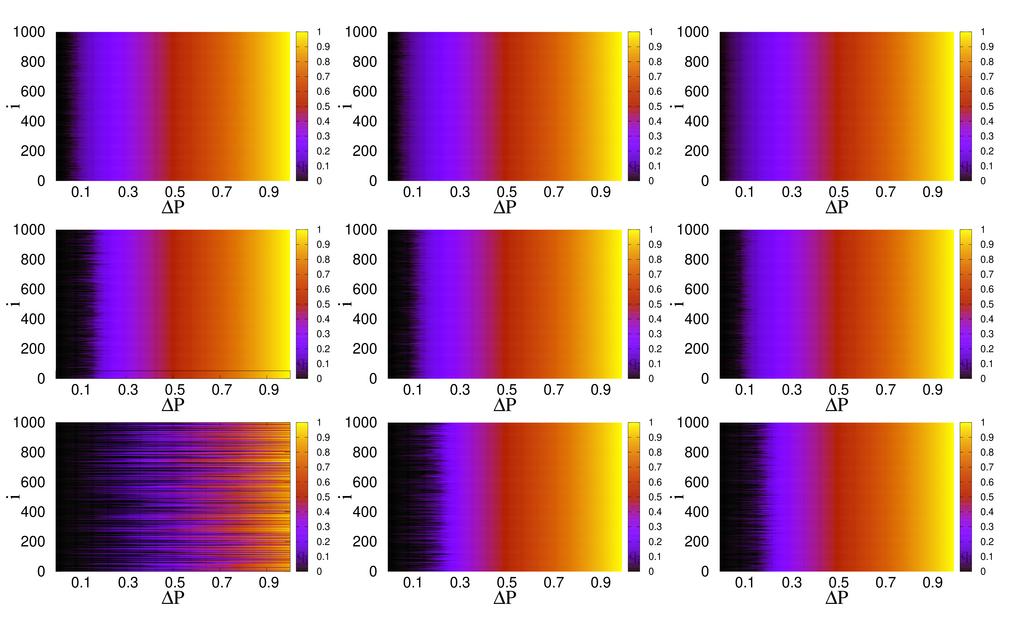}}
\caption{Local flow-rate shown with applied pressure along the x-axis and fiber index along y-axis. The figures are arranged as per increasing $k$ values: lower values lower panel, intermediate values middle panel and higher values upper panel. }
\label{fig8a}
\end{figure}
CFBM represents porous media as a bunch of tubes, where each tube referring to a pore link which acts as a flow path. Understanding individual flow paths may provide better idea to understand the  global nature. Local fluctuations in each flow path plays a major role in understanding the internal dynamics for a disordered system like CFBM.

Figure \ref{fig8a} shows the map of local flow at different applied pressure. Each subplot contains the flow-rate as a function of applied pressure $\Delta P$ (along x-axis) and fiber index $i$ (along y-axis). The lower panel corresponds to low $k$ values, the middle panel for intermediate $k$ values and the upper panel for high $k$ values. One can conclude visually from the map that as fracking amplitude increases the flow becomes more and more uniform. For low $k$, there is a high fluctuation in local flow rate even when the $\Delta P$ high suggesting strong localization where some tubes carry fluid at a much higher rate relative to others. The flow becomes more and more homogeneous as we increase fracking amplitude. Increasing fracking amplitude reduces the disorder by making thresholds comparable. This in turns, leads to rapid opening fluid paths at a smaller applied pressure. At sufficiently high pressure, all the active tubes carry almost similar flow rates, $\Delta P$ as $q_i \sim \sqrt{\Delta P^2-p_c^2}\approx \Delta P$, making flow profile homogeneous.

\begin{figure}[ht]
\centerline{\includegraphics[width=0.5\textwidth,clip]{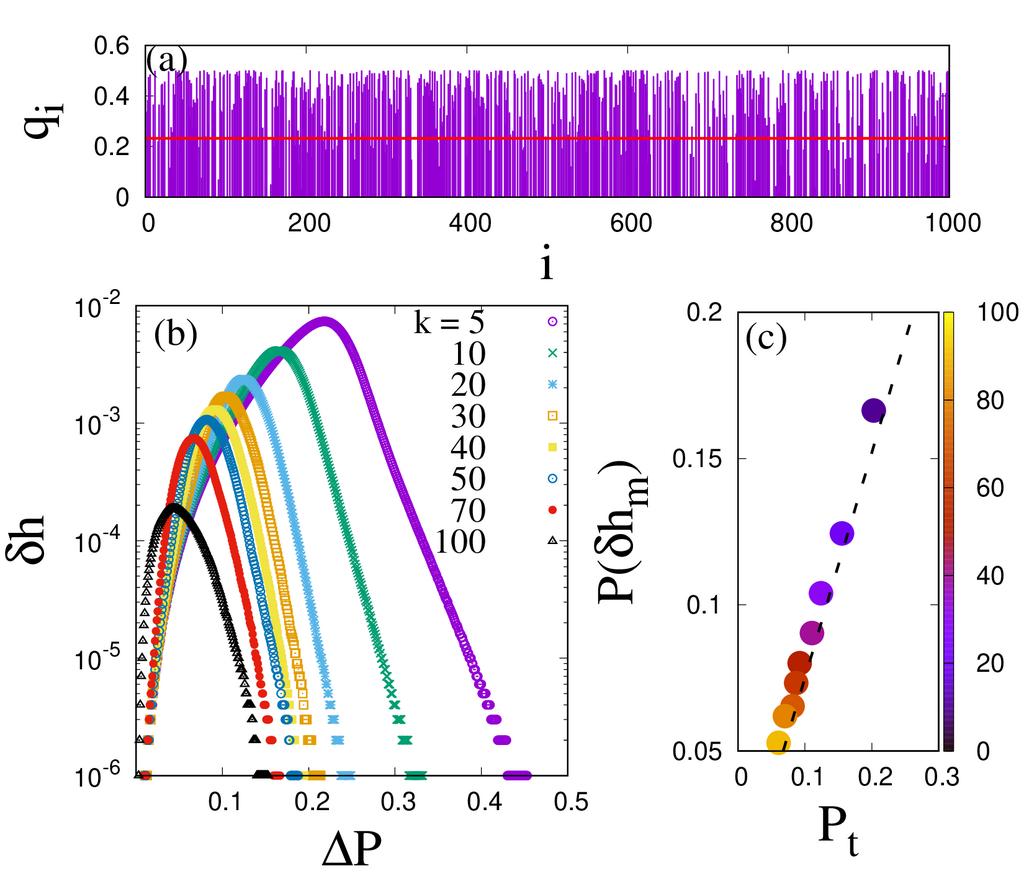}}
\caption{(a) Fluctuating flow rate as a function of tube index. The red line shows the spatial average for a single configuration. (b) Height difference $\delta h$ in flow profile as a function of $\Delta P$. (c) Following correlation between $P(\delta h_m)$ and $P_t$ is observed: $P_t \sim 1.43 P(\delta h_m)$.}
\label{fig9}
\end{figure}
To understand the heterogeneity in local flow profile in a better way, we have studied the fluctuation in flow rates from tube to tube about the average value for a single configuration. The idea is how much deviation does local flow in the each path shows with respect to the flow rate averaged over all tubes. Figure \ref{fig9}(a) shows the local flow values $q_i$ for a single configuration along with the corresponding tube index $i$. The red line is the average flow over all the tubes and given by: 
\begin{align}
\bar{q}=\displaystyle\frac{1}{N}\displaystyle\sum_{i=1}^N q_i
\end{align}
We want to stress on the fact that $\bar{q}$ is different from $<q>$ where the latter is obtained by taking another average of $\bar{q}$ over a number of configurations. We are not dealing with $<q>$  which is a global parameter as this section is strictly confined to local parameters only. The overall fluctuation for a profile like figure \ref{fig9}(a) is defined as 
\begin{align}
\delta h=\displaystyle\frac{1}{N}\displaystyle\sum_{i=1}^N (q_i-\bar{q})^2
\end{align}
Figure \ref{fig9}(b) shows the behavior of $\delta h$ as the pressure applied across the system is increased gradually. As expected, with increasing $\Delta P$, the model shows a non-monotonic fluctuation. For low pressure gradient, since very few tubes are open with similar capillary threshold, all flow profiles almost resembles each other, exhibiting strong channeling effect which reduces overall fluctuation. On the other hand, when $\Delta P$ is very high the flow rates are almost equivalent to $\Delta P$ due to the fact that $\Delta P>>p_c$ in that limit. For both very low and very high pressure, $\bar{q}$ is very close to individual $q_i$ values making the fluctuation $\delta h$ lower. For an intermediate $\Delta P$ the system undergo rapid opening of fluid paths, where the flow rate in each path deviates a lot from its average. As we increase fracking amplitude, the  spatial fluctuation reaches its maximum value $\delta h_m$ for an intermediate $\Delta P$. This maxima $\delta h_m$ is also observed to decrease with increasing $k$. As the fracking amplitude increases, the system undergoes reorganization due to the rapid opening of fluid pathways with increasing pressure but the threshold values shift towards zero much faster. This indeed reduces the fluctuation compared to a lower $k$ value and hence taking the maxima to a lower point. We have already seen that the global rheology shows a transition from nonlinear to linear behavior. This transition occurs just after the system reaches a state where the fluctuations begin to decrease, suggesting no more reorganization of fluid pathways. So, just before the onset of the Darcy flow, the fluctuation must be maximum. Figure \ref{fig9}(c) shows the following correlation between the maximum height fluctuation with the transition point $P_t$ from linear to non-linear region in the global rheology.
\begin{align}
P_t=1.24P(\delta h_m)-0.002
\end{align}
The above correlation predicts the transition point from the local fluctuation whose maxima takes places slightly before. This prediction plays an important role as transition point being a global parameter needs several configuration average, where as the local fluctuation being  single configuration data, the corelation between  this two local  and global parameters can reduce the computational cost heavily.
\begin{figure}[ht]
\centerline{\includegraphics[width=0.8\textwidth,clip]{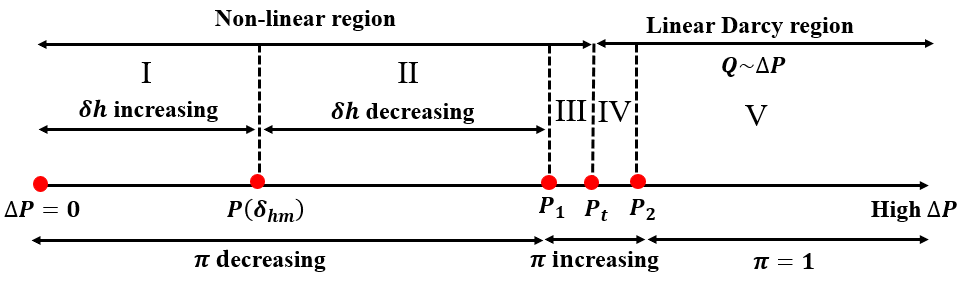}}
\caption{The timeline shows the the pressure values $P(\delta h_m)$, $P_1$, $P_2$ and $P_t$, especially which occurs after which sequentially. The correlation among above pressure values can be proven invaluable to bridge gap between local and global behavior.}
\label{fig_hf}
\end{figure}

We have observed that $P(\delta h_m)$ and $P_t$ are not the only two pressure values that can be correlated with each other. Earlier $P_1$ and $P_2$ is also observed to be correlated with $P_t$ and hence with $P(\delta h_m)$. The timeline in figure \ref{fig_hf} shows the sequence in which all this pressure values are observed. 
\begin{enumerate}[(a)]
    \item The first noticeable thing would be the maximum fluctuation in flow profile. So, $P(\delta h_m)$ takes place before any other pressure values.
    \item Next we observe the pressure $P_1$ at which the participation function ($\pi$) becomes minima. This is correlated with $P(\delta h_m)$ as: 
    \begin{align}
        P_1=1.13P(\delta h_m)-0.05
    \end{align}
    \item Next we come across the the onset of linear Darcy flow, which takes place at $\Delta P=P_t$. This correlation is just discussed above.
    \begin{align}
        P_t=1.24P(\delta h_m)-0.002
    \end{align}
    \item Finally when the pressure across the system is sufficiently high ($P_2$), $\pi$ becomes 1 suggesting equipartition of energy among all tubes. $P_2$ can also be correlated with $P(\delta h_m)$ as
    \begin{align}
        P_2=1.31P(\delta h_m)+0.04
    \end{align}
\end{enumerate}
This is important because of the very fact that $P_1$, $P_2$ and $P_t$ are global parameters that requires sufficient amount of configuration average. We will reduce the computation cost with appreciable amount if we can extract these global parameter from $P(\delta h_m)$ which requires only a single configuration. This will be even more effective for relatively complected model like Dynamic pore network that operated in higher dimension with computation cost increasing in a super-linear manner with size of the lattice. Below we have provided a table describing the features for the five different regions, $I$, $II$, $III$, $IV$ and $V$ shown in figure \ref{fig_hf}.  
\begin{center}
\begin{tabular}{||c | c | c ||} 
 \hline
 Region & Pressure & Characteristics \\ [0.5ex] 
 \hline
 $I$ & $\Delta P < P(\delta h_m)$ & \begin{minipage}[t]{8cm}
\begin{compactitem}
  \item $\Delta P-Q$ relation non-linear
  \item $\delta h$ increasing
  \item $\pi$ decreasing
\end{compactitem}
\end{minipage} \\
 \hline
 $II$ & $P(\delta h_m) \le \Delta P \le P_1$ & \begin{minipage}[t]{8cm}
\begin{compactitem}
  \item $\Delta P-Q$ relation non-linear
  \item $\delta h$ decreasing
  \item $\pi$ decreasing
\end{compactitem}
\end{minipage} \\
 \hline
 $III$ & $P_1 \le \Delta P \le P_t$ & \begin{minipage}[t]{8cm}
\begin{compactitem}
  \item $\Delta P-Q$ relation non-linear
  \item $\delta h$ decreasing
  \item $\pi$ increasing
\end{compactitem}
\end{minipage} \\
 \hline
 $IV$ & $P_t \le \Delta P \le P_2$ & \begin{minipage}[t]{8cm}
\begin{compactitem}
  \item $\Delta P-Q$ relation linear (Darcy flow)
  \item $\delta h$ decreasing
  \item $\pi$ increasing
\end{compactitem}
\end{minipage} \\
 \hline
 $V$ & $\Delta P > P_2$ & \begin{minipage}[t]{8cm}
\begin{compactitem}
  \item $\Delta P-Q$ relation linear (Darcy flow)
  \item $\delta h$ decreasing
  \item $\pi=1$ (equipartition of energy)
\end{compactitem}
\end{minipage} \\
 \hline
\end{tabular}
\end{center}


\subsection{Quantifying the extraction}
A true quantification of the fluid extracted will be the total amount of fluid collected when the pressure is gradually increased.  

\begin{figure}[ht]
\centerline{\includegraphics[width=0.5\textwidth,clip]{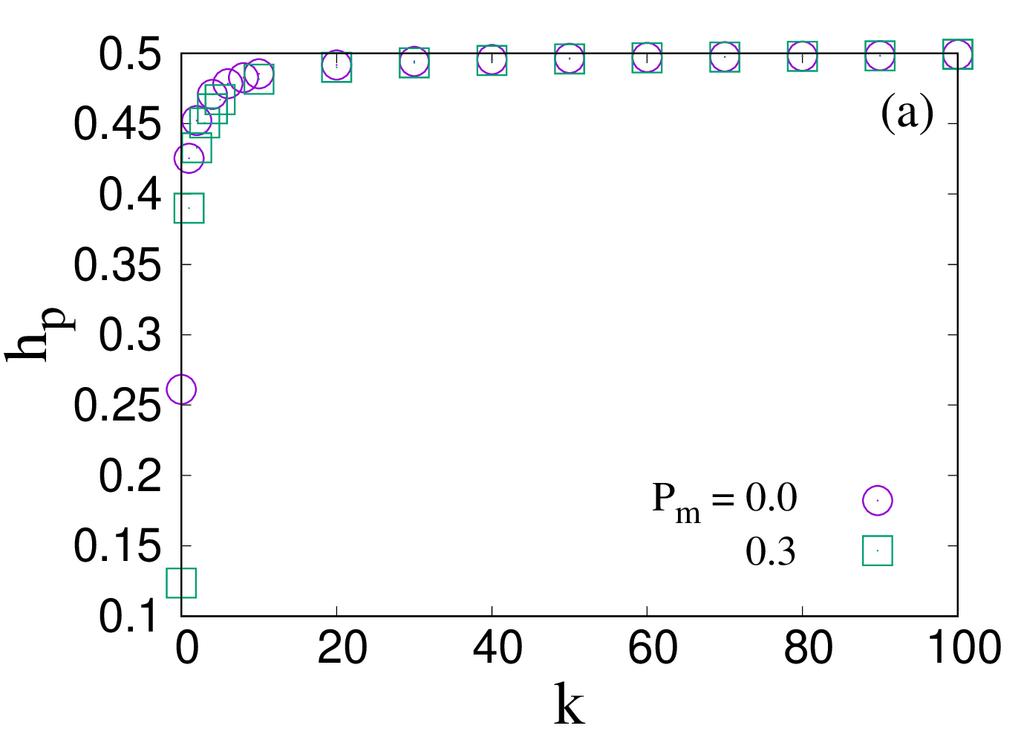}}
\centerline{\includegraphics[width=0.5\textwidth,clip]{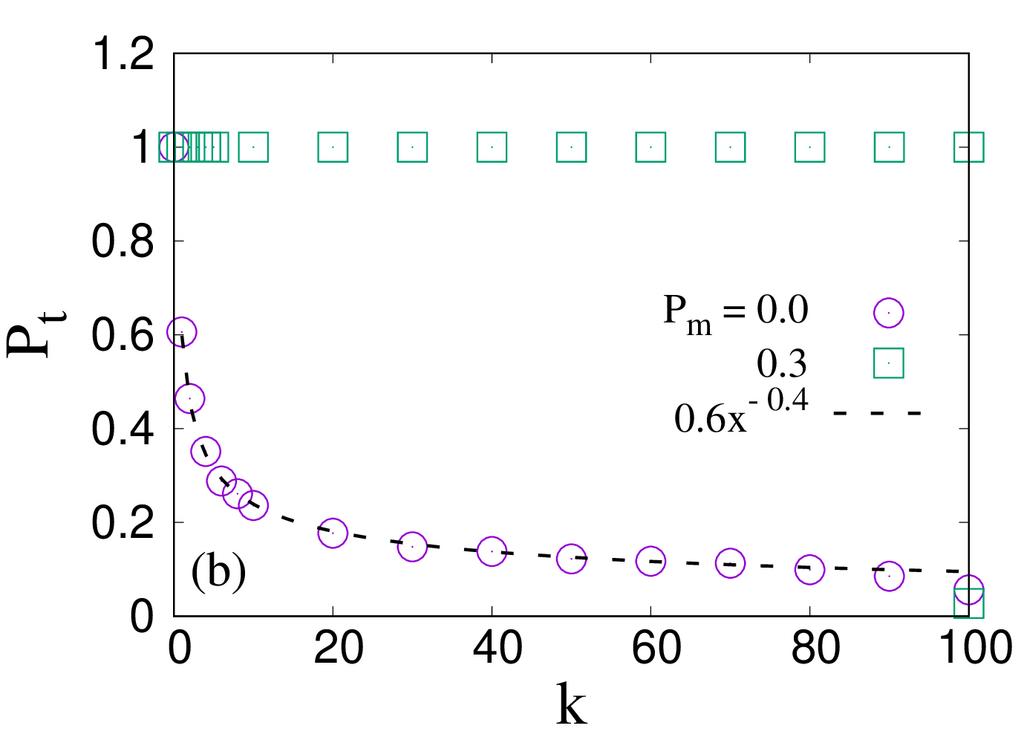}}
\caption{(a) The hydraulic power $h_p$ with $k$ for the range $0<\Delta P<1.0$, as beyond this the flow rate remains same irrespective of $k$. $h_p$ shows a sharp increment with $k$ for both $P_m=0$ and $P_m=0.3$, and remains constant around 0.5 for $k>8$ roughly. (b) For $P_m=0$, the transition point $P_t$ to the linear Dacry behavior falls sharply as $P_t \sim k^{-0.4}$. For $P_m=0.3$, $P_t$ remains constant at 1.0 independent of $k$ except for $k=0$ where it falls down to the same value as $P_m=0$.}
\label{fig4}
\end{figure}

Then the total fluid extracted in the limit $0<\Delta P<P_M(=1)$ will be given by the total area under $\Delta P-\langle q \rangle$ plot 
\begin{align}\label{eq12}
h_p=\displaystyle\int_{0}^{P_M}\langle q \rangle (\Delta P)d\Delta P=\displaystyle\int_{0}^{1}\langle q \rangle (\Delta P)d\Delta P
\end{align}
This is also known as the hydraulic power. The limit for $\Delta P$ is set between 0 to 1 as after 1 the rheology remains the same irrespective of the value of $k$. The only change in rheology due to the fracking event is observed before 1. We can use any numerical technique like Riemann sum, Trapezoidal rule, Simpsons 1/3 or 3/8 rule, etc to find $h_p$ and subsequent total extraction.

\begin{figure}[ht]
\centerline{\includegraphics[width=0.5\textwidth,clip]{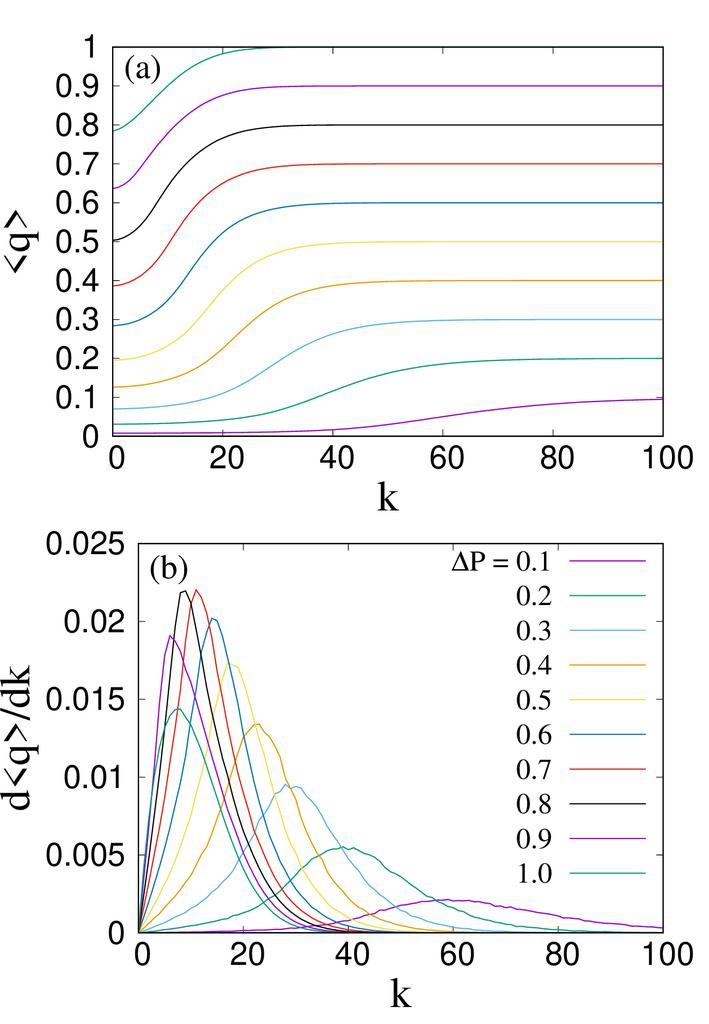}}
\caption{(a) Variation of average flow rate $\langle q \rangle$ with $k$ for $\Delta P$ ranging in between 0.1 and 1.0. (b) The relative change in $\langle q \rangle$ with respect to $k$. The rate of fluid flow with respect to $k$ is maximum at a certain pressure gradient $P^{\ast}$. The hydraulic fracture is most effective at this pressure.}
\label{fig5}
\end{figure}

Figure \ref{fig4}(a) shows how $h_p$ behaves at both $P_m=0$ and $0.3$. In both cases, $h_p$ increases rapidly with $k$ and saturates at 0.5 for $k>8$ roughly. This increment is due to the fracking events which have a prominent role in the total volume of extracted fluid. The non-linear to Darcy like transition point $P_t$, on the other hand, shows different behavior depending on whether $P_m$ is zero or not. For $P_m=0$, $P_t$ drops in a scale-free manner with $k$: $P_t \sim k^{-0.4}$, and saturates around 0.1, which further drops to a lower value for $k=100$. For $P_m=0.3$, $P_t$ remains almost constant at 1.0 unless $k=100$ where $P_t$ drops to a very low value.   

\begin{figure}[ht]
\centerline{\includegraphics[width=0.5\textwidth,clip]{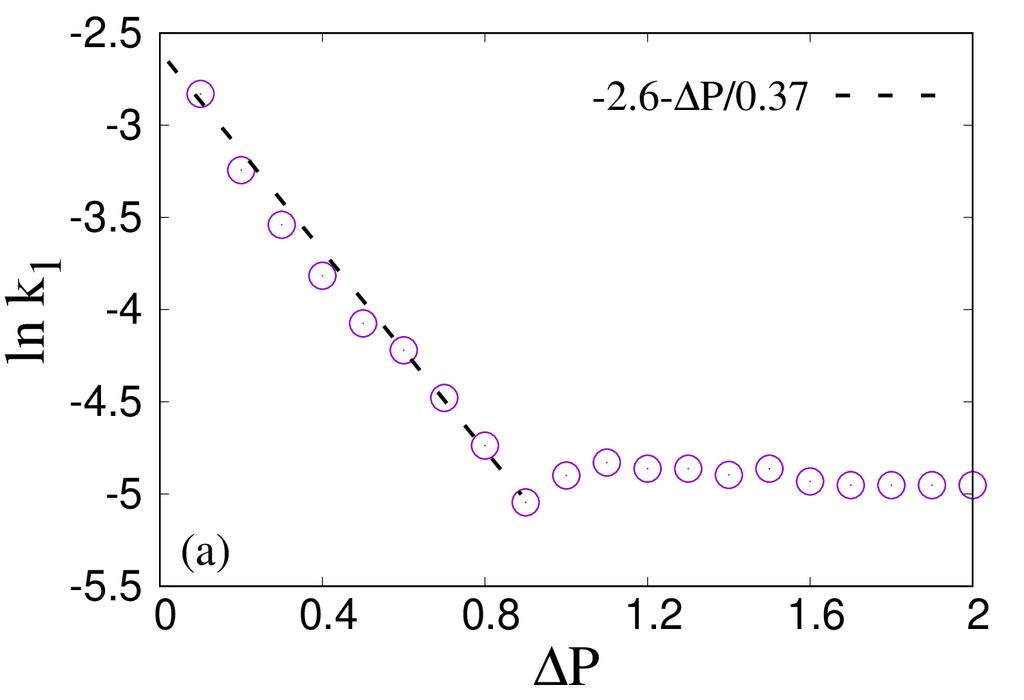}}
\centerline{\includegraphics[width=0.5\textwidth,clip]{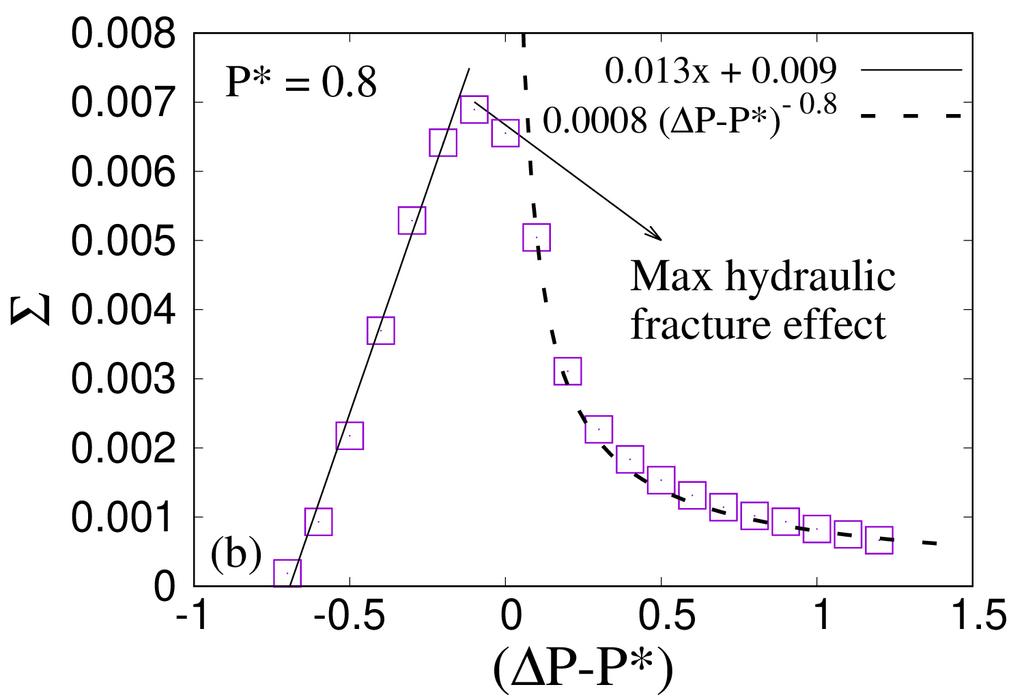}}
\caption{(a) Value of $k=k_1$ at which one observes the maximum rate of fluid extraction $d\langle q \rangle/dk$ as a function of $\Delta P$. $k_1$ continuously drops until $\Delta P=0.8$ and remains constant afterwards. (b) Variation of the $\Sigma$, the effectiveness of hydraulic fracture as a function of applied pressure. $\Sigma$ shows a non-monotonic behavior with peak at $\Delta P=P^{\ast}=0.8$. For $\Delta P<P^{\ast}$, $\Sigma \sim \Delta P$, while for $\Delta P<P^{\ast}$, we observe $\Sigma \sim (\Delta P-P^{\ast})^{-0.8}$. $P^{\ast}$ is the applied pressure where the hydraulic fracture makes the most impact.}
\label{fig6}
\end{figure}

A rather bigger picture can be understood by observing closely how the flow rate $\langle q \rangle$ changes with $k$ while keeping the pressure gradient $\Delta P$ constant. Figure \ref{fig5}(a) shows how $\langle q \rangle$ increases with $k$ and finally reaches a steady state value at high $k$. The point after which $\langle q \rangle$ becomes constant depends on $\Delta P$ so does the rate at which the flow rate increases. Figure \ref{fig5}(b) explicitly shows the rate of change of flow rate with $k$, $d\langle q \rangle/dk$, that shows a non-monotonic behavior for a constant $\Delta P$ showing a peak at $k_1$ with magnitude $dq_{max}$. We observe $dq_{max}$ increases gradually with $\Delta P$, reach a maxima at $\Delta P\approx0.8$ and decreases afterwards.   

\begin{figure}[ht]
\centerline{\includegraphics[width=0.5\textwidth,clip]{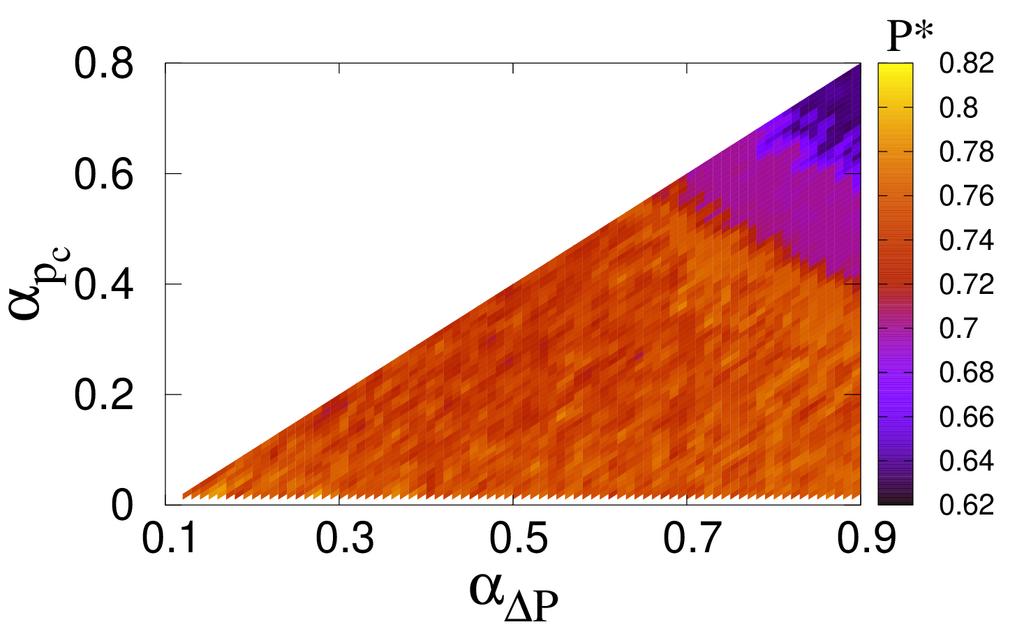}}
\caption{Variation of the optimum pressure $P^{\ast}$ as functions of $\alpha_{\Delta P}$ and $\alpha_{p_c}$. We always keep $\alpha_{\Delta P}>\alpha_{p_c}$ as $\Delta P$ is more dominant for fracking. We observe a decrease in $P^{\ast}$ with with $\alpha_{p_c}$ when $\alpha_{\Delta P}$ is high. For low $\alpha_{\Delta P}$, $\Sigma$ remains almost constant irrespective of $\alpha_{p_c}$.}
\label{fig7}
\end{figure}

\subsubsection{Extraction efficiency ($\Sigma$)} 
At this point, we can define a efficiency of fluid extraction, $\Sigma$, through the quantities we already have explored. The main idea of fracking is to figure out the condition for which maximum change in fluid extraction takes place with a larger rate. Figure \ref{fig6}(a) shows that $k_1$ decreases with $\Delta P$ suggesting we need less fracking amplitude when the pressure gradient is sufficiently high. Though we don't need to keep increasing $\Delta P$ as $k_1$ remains constant after 0.8 and increasing $\Delta P$ further will not decrease $k_1$ further but might increase the tendency of induced seismicity in real systems at higher pressure gradient. We define the efficiency of the fluid extraction by $\Sigma$ (see figure \ref{fig6}b) which is the product of $dq_{max}$ with $\Delta q$ the total change in $\langle q \rangle$ due to the change in $k$. 
\begin{align}
\Sigma=dq_m \times \Delta q 
\end{align}
$\Delta q$ is the absolute difference between the flow rate at $k=0$ and its steady state value at high $k$. The maximum $\Sigma$ stands for the scenario where maximum change ($\Delta q$) in extracted fluid takes place with a faster rate ($dq_m$). We observe $\Sigma$ to be maximum at $\Delta P=P^{\ast}(=0.8)$ and falls down on either side of it. At lower pressure, it increases linearly and falls down in a scale free manner beyond $P^{\ast}$.
\begin{align}
\Sigma &= 0.013(\Delta P-P^{\ast})+0.0009 \ \text{for} \ \Delta P<P^{\ast} \ \text{and} \nonumber \\  
       &= 0.0008(\Delta P-P^{\ast})^{-0.8}  \ \text{for} \ \Delta P>P^{\ast}
\end{align}
Finally, for the completeness of our study, we have observed the variation of $P^{\ast}$ when both the parameters $\alpha_{\Delta P}$ and $\alpha_{p_c}$ simultaneously varied (see figure \ref{fig7}). $\alpha_{\Delta P}$ ranges from 0.1 to 0.9 while $\alpha_{p_c}$ from 0.0 to 0.8. This satisfies our criteria $\alpha_{\Delta P}>\alpha_{p_c}$. For higher $\alpha_{\Delta P}$, $P^{\ast}$ starts from 0.8 at low $\alpha_{p_c}$ and decreases to 0.68 when $\alpha_{p_c}$ is high. At the same time, $P^{\ast}$ shows a slight decrease as we $\alpha_{\Delta P}$ while keeping $\alpha_{p_c}$ fixed.  


\subsubsection{Universality in extraction}
To check the universality of our results, we have repeated the study with 3 other threshold distributions, gaussian, power law, and Weibull for the capillary barrier other than when they are uniformly distributed.

\begin{equation}
\label{eqn4-10}
\rho(p_c)=\left\{\begin{array}{ll}
                                0       & \mbox{, $p_c \le P_m$\;,}\\
                                \displaystyle\frac{1}{\sigma\sqrt{2\pi}}e^{-\frac{1}{2}(\frac{p_c-\mu}{\sigma})^2} & \mbox{, $P_m < p_c \le P_M$\:,}\\
                                0       & \mbox{, $p_c > P_M$\;,}\\
              \end{array}
       \right.
\end{equation}
\begin{equation}
\label{eqn4-10}
\rho(p_c)=\left\{\begin{array}{ll}
                                0       & \mbox{, $p_c \le P_m$\;,}\\
                                p_c^{\alpha} & \mbox{, $P_m < p_c \le P_M$\:,}\\
                                0       & \mbox{, $p_c > P_M$\;,}\\
              \end{array}
       \right.
\end{equation}
\begin{equation}
\label{eqn4-10}
P(p_c)=\left\{\begin{array}{ll}
                                0       & \mbox{, $p_c \le P_m$\;,}\\
                                \displaystyle\frac{k}{\lambda}\left(\displaystyle\frac{p_c}{\lambda}\right)^{k-1}e^{-(p_c/\lambda)^k} & \mbox{, $P_m < p_c \le P_M$\:,}\\
                                0       & \mbox{, $p_c > P_M$\;,}\\
              \end{array}
       \right.
\end{equation}
To keep the same permeability, we have set the parameters in the following way for the distributions: (i) mean $\mu=0.5$, and variance $\sigma=0.5$ for gaussian, (ii) slope $\alpha=-1$ for power law, and (iii) shape parameter $k=1.2$, and scale parameter $\lambda=1.2$ for Weibull.
\begin{figure}[ht]
\centerline{\includegraphics[width=0.5\textwidth,clip]{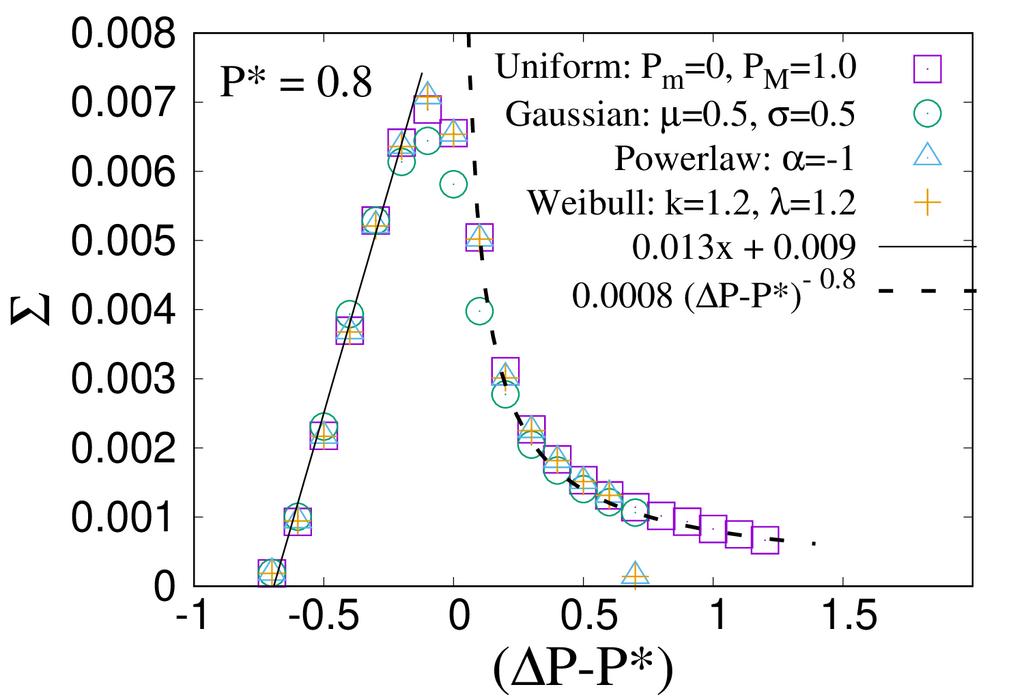}}
\caption{Variation of the $\Sigma$, the extractivity during hydraulic fracture as a function of applied pressure for all four threshold distributions - uniform, gaussian, Weibull, and power law. $P^{\ast}$ and the nature of $\Sigma$ before and after $P^{\ast}$ is not affected by the choice of the distribution if the permeability is kept uniform throughout all distributions.}
\label{fig8}
\end{figure}
Figure \ref{fig8} shows the variation of effectiveness $\Sigma$ with applied pressure $\Delta P$ for all four distributions considered to check universality. The optimum pressure gradient $P^{\ast}$ is observed to be independent of the choice of the distribution providing the mobility remains unaltered. The linear increase of $\Sigma$ before $P^{\ast}$ and the scale-free fall afterwards also remains unaltered irrespective of the choice of the distribution.

\subsubsection{Entropy of local flow configuration}
To capture prominently how the local flow dynamics change with minute change in the system,  we have introduced the concept of entropy here. The idea of Shannon entropy is to  captures the local re-organization of flow through a single parameter. We wanted to see how the fluctuation differences in individual local flows contribute to the global phenomena and whether it can be used to predict the pressure corresponding to the optimum extractivity. 
\begin{figure}[ht]
\centerline{\includegraphics[width=0.5\textwidth,clip]{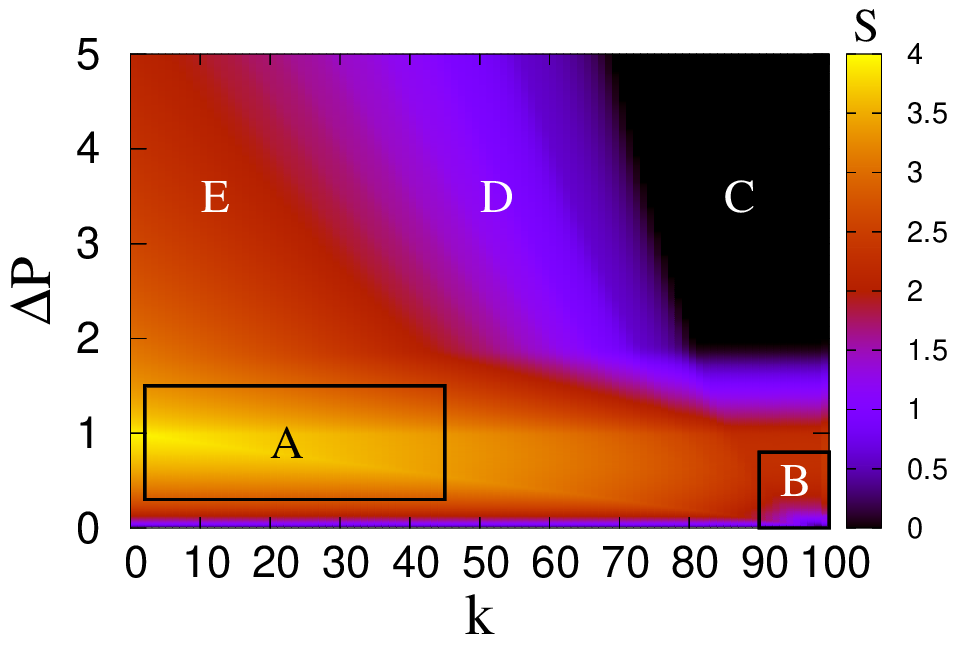}}
\caption{Entropy $S$ as functions of both fracking amplitude $k$ and applied pressure $\Delta P$. Depending on how $S$ behaves, the whole $k-\Delta P$ plane is divided among five regions: A $-$ non-monotonic behavior in $S$ which is very sharp for low $k$ but spreads for higher $k$; B $-$ sudden increase in $S$ for $k=100$; C $-$ zero entropy region; D $-$ $S$ decreasing at faster rate with $k$; and E $-$ $S$ decreasing at a relatively slower rate with $k$.}
\label{fig13}
\end{figure}

The Shanon entropy for a certain flow configuaration, obtained for a specific mobility and fracking amplitude is given as, 
\begin{align}
S=-\displaystyle\sum_{q_w}P(q_w)\ln[P(q_w)]
\end{align}
where $P(q_w)$ is the probability of finding a tube with local flow rate ranging between $q_w$ and $q_w-w$, $w$ is the size of the window for calculating the frequencies of local $q$ values. $P(q_w)$ is calculated by dividing the count of tubes for a certain window by the {\it total count} for that fluctuating series. $S_d$ is the dynamic entropy which keeps changing (increasing) with $q_w$. We get the total entropy $S$ by summing over all possible $q_w$ values within that fluctuating series.  
\begin{figure}[ht]
\centerline{\includegraphics[width=0.5\textwidth,clip]{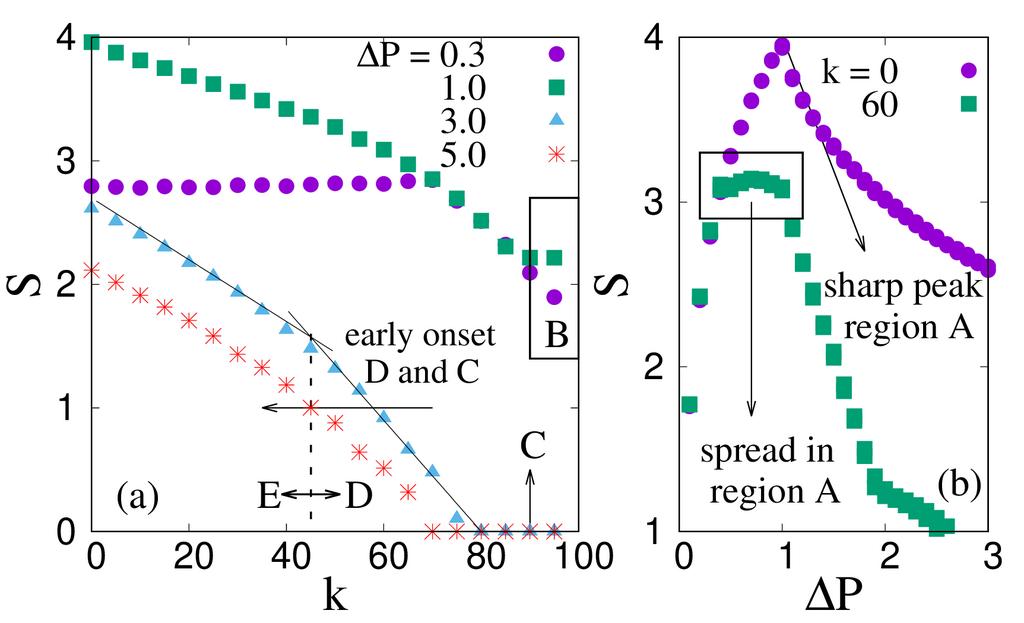}}
\caption{Variation of the entropy $S$ with both $k$ and $\Delta P$ respectively keeping the other parameter constant. The figure shows the five different regions, A to E, described in figure \ref{fig13}.}
\label{fig14}
\end{figure}

Figure \ref{eq13} gives a better idea about the entropy when it is plotted simultaneously with both $k$ and $\Delta P$. The color gradient is placed for $S$ which varies from 0 to 4. Depending on $S$ the whole $k-\Delta P$ plane is divided in five different regions. We will discuss the regions one by one below. We will discuss both figure \ref{fig14} and figure \ref{fig13} together while discussing these different regions.\\
{\bf Region A -} This region shows a non-monotonic behavior of $S$ with $\Delta P$. This happens due to the fact that fluctuation in $q_i$ is lesser for both very low and and very high pressure gradient, leading to lower entropy in those limiting cases. At any an intermediate $\Delta P$, the fluctuation as well as the entropy is high. For low $k$ this non-monotonic behavior shows a distinct peak while for higher $k$ this maxima exists for a range of $\Delta P$ evident from the spread of region $A$ when $k$ is high. Figure \ref{fig14}(b) shows this non-monotonic behavior for a both high and low value of $k$. For $k=0$, we can see the sharp peak while for $k=60$ we see the maximum over a range, evident by the spread of the region A (yellow color) in figure \ref{fig13} as well.\\
{\bf Region B -} This region shows a slight increase in $S$ for $k=100$. $S$ decreases with $k$ due to reduced fluctuation but then again shows a slight higher value at region B ($k=100$) since the spread of local flow rates remains unchanged in this case, only some of the threshold values becomes zero. This is shown explicitly in figure \ref{fig14}(a) for $\Delta P=0.3$, and 1.0.\\
{\bf Region C -} This region corresponds to $S=0$. The entropy of the system will be close to zero if the fluctuation in the system is very small. This happens at high $k$ and high $\Delta P$. In this limit $\Delta P>>p_c$, producing a flow rate $q_i \sim \sqrt{\Delta P^2-p_c^2} \sim \Delta P$, irrespective of the individual $p_c$ values. The region shows that a higher $\Delta P$ is required to achieve $S=0$ for relatively lower $k$. This region is also shown in figure \ref{fig14}(a) where onset of region C is pointed by the starting point of $S=0$, which shifts to lower values as $\Delta P$ increases. \\
{\bf Region D \& E -} In both these regions, the entropy decreases with $k$ but the rate of decrease is faster in region $E$. This is shown in figure \ref{fig14}(a) for $\Delta P=3.0$, through two straight lines with different slopes and the boundary of D and E can be approximated by the intersection of these two straight lines (shown by a vertical dotted line).

The study of entropy with respect to both pressure and fracking amplitude measures how evenly the total flow is shared among the available pathways and how the system reorganizes itself when exposed to increment in pressure and fracking amplitude.


\subsection{Prediction of maximum extractivity from Shanon Entropy: the role of mobility}
\begin{figure}[ht]
\centerline{\includegraphics[width=0.5\textwidth,clip]{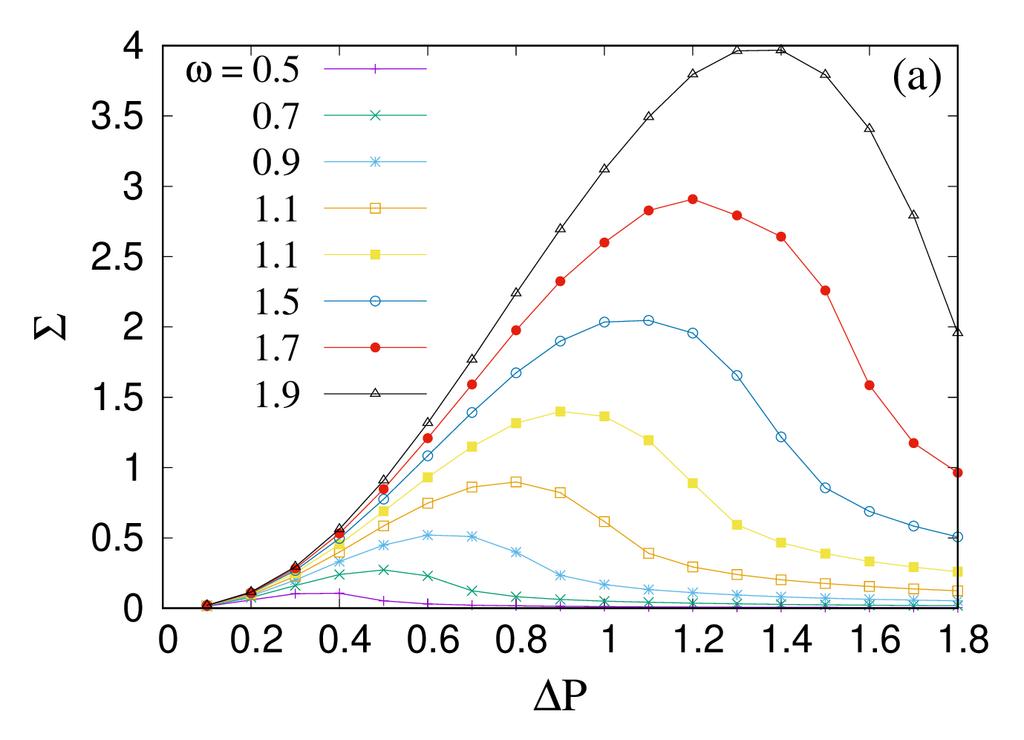} \ \includegraphics[width=0.5\textwidth,clip]{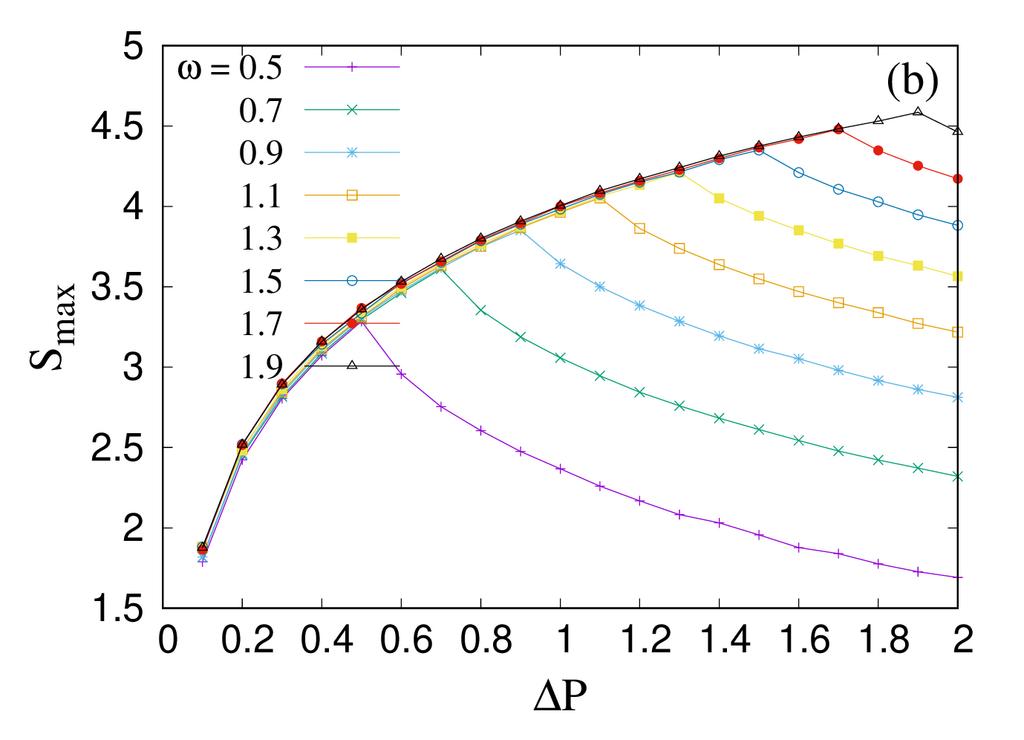}}
\caption{(a) Variation of extractivity $\Sigma$ with $\Delta P$ for disorder width varying between 0.5 and 1.9. $\Sigma$ shows a peak at $P^{\ast}$ which moves to higher values as we increase $\omega$. (b) Variation of the maximum entropy $S_{max}$ with $\Delta P$ for $0.5\le\omega\le1.9$. $S$ also shows a non-monotonic behavior with maxima at $P_S$, which also scales to higher values with $\omega$.}
\label{fig16}
\end{figure}
The final thrust to our work is to correlate the behavior of entropy extracted from the local flow profile with optimum extractivity that we observed earlier. For a particular mobility, set by the width of the distribution of capillary threshold, we will get a single value of $\Sigma_m$ as well as entropy. To validate the correlation, we have tune the width in order to attain different mobility and monitor how above parameters behave with that variation. The width is inversely proportional to the mobility of the system - higher the width lower the mobility as we are including tubes with higher capillary pressure and hence low or no flow associated with it even when the applied pressure is high. Consequently, the problem is closely related to how flow paths become activated, depending on the availability and distribution of accessible channels within the porous medium. Figure \ref{fig16}(a) shows the variation of $\Sigma$ with increasing pressure gradient $\Delta P$ while the distribution width is varied from $0.5$ to $1.9$. The maximum value, $\Sigma_m$, moves to higher values as we increase the width $\omega$ of the distribution. The pressure $P^{\ast}$, at which this maxima occurs also shifts towards higher values. A similar trend is observed for the entropy $S$ produced at a particular mobility while increasing the pressure gradient. $S$ also increases with $\Delta P$, reaches a maxima $S_{max}$ at $\Delta P=P_S$ and then decreases. $P_S$ is also observed to move to higher values as we increase $\omega$ decreasing the effective mobility of the system. 
\begin{figure}[ht]
\centerline{\includegraphics[width=0.5\textwidth,clip]{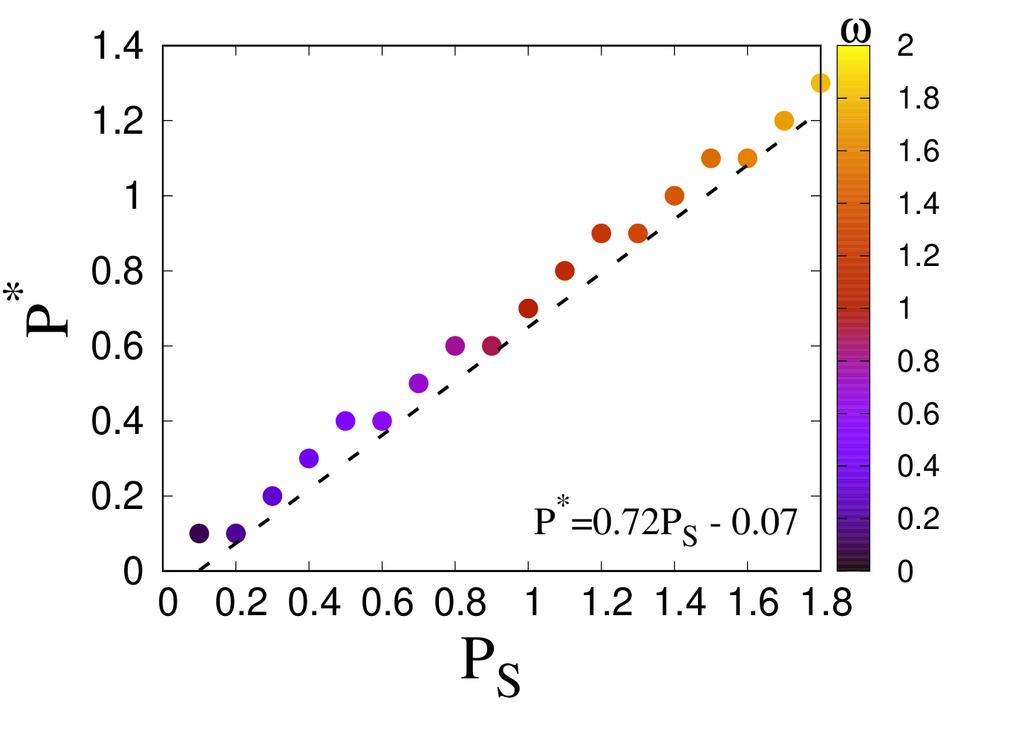}}
\caption{Correlation between $P^{\ast}$ and $P_S$ as the disorder width is tuned between 0.2 and 2.0. We observe the following linear correlation: $P^{\ast}=0.72P_S-0.07$.}
\label{fig17}
\end{figure}

Figure \ref{fig17} shows the correlation between $P_S$ and $P^{\ast}$. We observe the following behavior, 
\begin{align}
P^{\ast}=0.72P_S-0.07
\end{align}
Though the correlation is very strong here, the only problem is $P^{\ast}$ takes place after $P_S$, which means as we increase the applied pressure, the extractivity becomes maximum first and then we reach the maximum entropy scenario. This will not be helpful in predicting $P^{\ast}$ from the knowledge of $P_S$. To take care of this scenario we next have calculated the relative change in entropy with respect to $k$, $dS/dk$, representing the rapid change in flow profile. 
\begin{figure}[ht]
\centerline{\includegraphics[width=0.5\textwidth,clip]{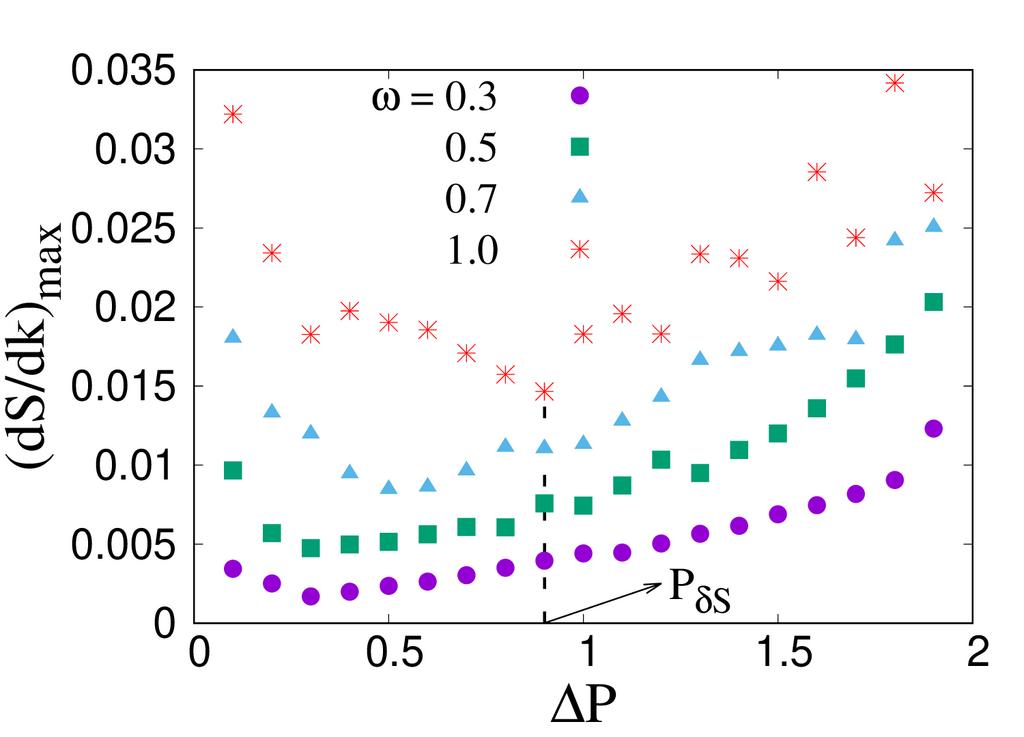}}
\caption{Variation of $|dS/dk|_{max}$ with applied pressure $\Delta P$. $|dS/dk|_{max}$ shows a minima at $\Delta P=P_{\delta S}$ and increases as we move to higher or lower pressure gradient.}
\label{fig18}
\end{figure}

The study is repeated for different mobility and applied pressure values. Figure\ref{fig18} shows the behavior of  $|dS/dk|_{max}$, the maximum of relative change in $S$, with $\Delta P$ for different width ($\omega$) values. One can clearly observe from the trend that $|dS/dk|_{max}$ shows a minima at a certain applied pressure, say $P_{\delta S}$. $P_{\delta s}$ refers to the point where the system is undergoing a rapid change but at a slower rate compared to other $\Delta P$. An extraction will be more efficient if the flow profile inside the porous media shows less local fluctuation. We believe this point will be connected closely with $\Sigma_m$. 
\begin{figure}[ht]
\centerline{\includegraphics[width=0.5\textwidth,clip]{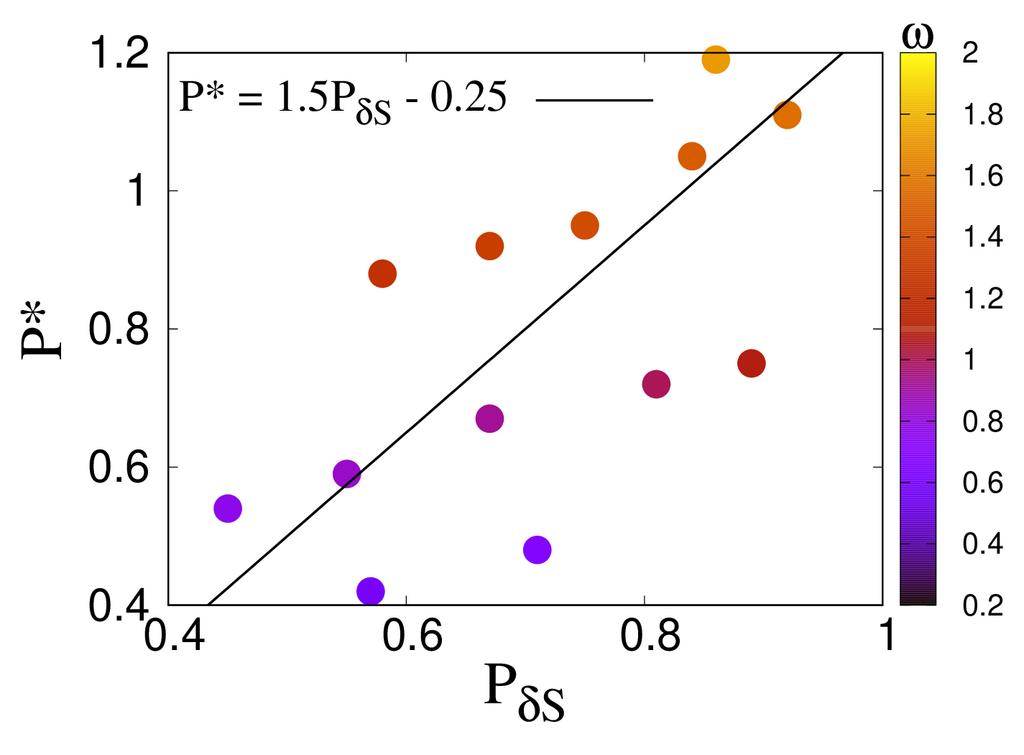}}
\caption{Correlation between $P^{\ast}$ and $P_{\delta S}$ as the disorder width is tuned between 0.2 and 2.0. We observe the following linear correlation: $P^{\ast}=1.5P_{\delta S}-0.25$.}
\label{fig19}
\end{figure}

Figure \ref{fig19} shows the following correlation of $P^{\ast}$ with $P_{\delta S}$ with varying $\omega$: 
\begin{align}
P^{\ast}=1.5P_{\delta S}-0.25
\end{align}
This correlation is extremely useful for us since here $P^{\ast}$ takes place after $P_{\delta S}$. This means we observe the minima of $|dS/dk|_{max}$ at a lower pressure gradient and we can use the above correlation to estimate $P^{\ast}$ and make our prediction complete. This correlation is stronger for lower $\omega$ and becomes weaker when $\omega$ is high. At higher $\omega$, there is enough fluctuation already present in the system and finding a proper correlation like this is relatively harder. 


\section{Conclusion}
The problem of hydraulic fracture have a fair share within the fields of multi-phase flow, mechanical heterogeneity of core body, fluid dynamics and fracture mechanics. In the literature, such problems are handled through indispensable tools like statistical modeling (for both fracture and flow) - ranging from analytical treatment to high-fidelity numerical simulations. Models like linear elastic fracture mechanics \cite{irwin57}, field-scale simulations \cite{lb01}, discrete fracture network (DFN) \cite{hljc21}, particle-based methods \cite{zzbh16} has been involved to understand the fracture nature. The models being complicated enough as it is could be burdened with the extra complexity of fluid flow through fractured pores, especially in presence of heterogeneity and anisotropy. The above mentioned methods, though extremely capable to represent failure process in disordered media and subsequent fluid flow in the fractured layers, the tradeoff, however, is the high computational cost due to the complexity of the models, especially in higher dimension and with higher system sizes.  

In the paper, without going into much complexity, we discuss through capillary fiber bundle model, a detailed quantification fluid extraction efficiency and subsequent change in rheology as we include the hydraulic fracture events during a multi-phase flow through a porous media. The maximum rate of change in $\langle q \rangle$ with respect to $k$, combined with the absolute change in the flow rate ($\Delta q$), gives the efficiency of fluid extraction, denoted by $\Sigma$ and defined as: $dq_m \times \Delta q$. The efficiency $\Sigma$ becomes maximum at a intermediate pressure gradient $P^{\ast}$ and not when $\Delta P$ is very high. This suggests that one does not need to go to very high magnitude $k$ of fracking or very high pressure to achieve optimum effect of the hydraulic fracture event.

Further, the dynamics of hydraulic fracture depends closely on how a fracture induced flow path effect the global dynamics as a whole. To understand the local flow better, we have done a systematic study of the role of local fluctuation in each pathway. Fluctuation in a fluid pathway refers to, how much deviation does  the flow rate in each pathway shows compare to average flow rate. The non- monotonic behavior of fluctuation shows how system reorganizes itself with increment in pressure and fracking amplitude. The fluctuations are further studied deeply by calculating the participation number which quantifies how the flow rates are distributed across the available flow paths within the system. The participation number in power spectrum predicting change in noise i.e, from red to black is a sign that high fracture amplitude can trigger seismic activity.

As local behavior provides valuable insights into underlying dynamics, the key role of studying this is to understand how it predicts the global behavior and the subsequent correlation between them. The maximum fluctuation is correlated with the global onset of Darcy flow. This prediction can highly reduces the computational cost as we can predict the point of transition from a single sampled data of local flow. At the same time entropy being the parameter that captures the reorganization of fluid pathways is correlation with maximum extractivity. This is done by tuning the mobility through the system controlled by the witdth of capillary threshold distribution. These valuable correlations connect the local and global behavior of the system focusing on how even a single pathway triggered by fracking has an impact on whole system dynamics. 

In conclusion, a comprehensive understanding of hydraulic fracturing requires continued integration of physics-based models with data-driven techniques, guided by experimental observations and validated against field-scale outcomes. By advancing the physical realism, computational efficiency, and interpretability of fracture models, future research will be well-positioned to support safer, more effective, and more sustainable hydraulic fracturing practices. A natural extension to this work will be incorporating the fracking events in two dimension in dynamic pore network model (DPNM) \cite{sgvh20}, which is explored extensively in the context of multiphase flow to understand the rheology \cite{rsh20}, role of porosity \cite{rsh21}, role of wetting condition \cite{fsrh21}, nature of transient behavior \cite{smfrh24}, as well as introducing a new thermodynamic approach \cite{rpsh22}. The element of fracking in CFBM can complement the path-opening dynamics and hence shed light on the complex rheology, a topic which is recently explored by the same authors \cite{vr25}. The robust nature of the model makes it ideal for exploring the fracking events under the influence of different external parameter. This will allow us not only observe the steady-state but also go through the transient behavior, the latter of which will be extremely relevant in the context of hydraulic fracture.     

\section{Acknowledgment}
This work is supported by ANRF grant number SRG/2023/000866. 
\bigskip


\end{document}